\documentclass[%
 reprint,
 superscriptaddress,
nofootinbib,
 amsmath,amssymb,
 aps,
 prd,
 floatfix
]{revtex4-1}

\usepackage{graphicx}
\usepackage{dcolumn}
\usepackage{bm}
\usepackage{amssymb}
\usepackage{color}
\usepackage{multirow}
\usepackage[caption=false]{subfig}

\def\fh{{\it FrequencyHough}}

\newcommand{\Fstat}{$\mathcal{F}$-statistic}
\newcommand{\F}{\mathcal{F}} 
\newcommand{\EatH}{Einstein@Home} 
\def\td{{\it Time-Domain \Fstat}}

\begin{document}

\preprint{APS/123-QED}

\title{All-sky search for continuous gravitational waves from isolated neutron stars \\ using Advanced LIGO O2 data}

\author{B.~P.~Abbott}
\affiliation{LIGO, California Institute of Technology, Pasadena, CA 91125, USA}
\author{R.~Abbott}
\affiliation{LIGO, California Institute of Technology, Pasadena, CA 91125, USA}
\author{T.~D.~Abbott}
\affiliation{Louisiana State University, Baton Rouge, LA 70803, USA}
\author{S.~Abraham}
\affiliation{Inter-University Centre for Astronomy and Astrophysics, Pune 411007, India}
\author{F.~Acernese}
\affiliation{Universit\`a di Salerno, Fisciano, I-84084 Salerno, Italy}
\affiliation{INFN, Sezione di Napoli, Complesso Universitario di Monte S.Angelo, I-80126 Napoli, Italy}
\author{K.~Ackley}
\affiliation{OzGrav, School of Physics \& Astronomy, Monash University, Clayton 3800, Victoria, Australia}
\author{C.~Adams}
\affiliation{LIGO Livingston Observatory, Livingston, LA 70754, USA}
\author{R.~X.~Adhikari}
\affiliation{LIGO, California Institute of Technology, Pasadena, CA 91125, USA}
\author{V.~B.~Adya}
\affiliation{Max Planck Institute for Gravitational Physics (Albert Einstein Institute), D-30167 Hannover, Germany}
\affiliation{Leibniz Universit\"at Hannover, D-30167 Hannover, Germany}
\author{C.~Affeldt}
\affiliation{Max Planck Institute for Gravitational Physics (Albert Einstein Institute), D-30167 Hannover, Germany}
\affiliation{Leibniz Universit\"at Hannover, D-30167 Hannover, Germany}
% ...

\collaboration{The LIGO Scientific Collaboration and the Virgo Collaboration}

\author{A.~Pisarski}
\affiliation{University of Bia{\l }ystok, 15-424 Bia{\l }ystok, Poland}

\begin{abstract}
We present results of an all-sky search for continuous gravitational waves (CWs), which can be produced by fast spinning neutron stars with an asymmetry around their rotation axis, using data from the second observing run of the Advanced LIGO detectors. Three different semi-coherent methods are used to search in a gravitational-wave frequency band from 20 to 1922 Hz and a first frequency derivative from $-1\times10^{-8}$ to $2\times10^{-9}$ Hz/s. None of these searches has found clear evidence for a CW signal, so upper limits on the gravitational-wave strain amplitude are calculated, which for this broad range in parameter space are the most sensitive ever achieved.
\end{abstract}

\maketitle

\section{Introduction}

Eleven detections of gravitational waves from black hole binaries and from a neutron star binary have been reported in \cite{Catalog}. One of the characteristics of the signals detected so far is that their duration ranges from a fraction of second to tens of seconds in the detector sensitive frequency band. Other mechanisms, however, can produce gravitational waves with longer durations, not yet detected. In this paper we describe an all-sky search for continuous gravitational waves (CWs), almost monochromatic signals which are present at the detectors during all the observing time. The principal sources for CW emission (see \cite{CWReview} for a review) are spinning neutron stars. If a spinning neutron star (NS) has an asymmetry with respect to its rotation axis, it will emit CWs at twice the rotation frequency. 

Fast-spinning neutron stars in the Milky Way can generate continuous gravitational waves via various processes which produce an asymmetry. Crustal distortions from cooling or from binary accretion, or magnetic field energy buried below the crust could lead to the non-axisymmetry necessary for detectable emission. The excitation of r-modes in a newborn or accreting NS is another promising mechanism for the emission of CWs. Recently, some evidence for a limiting minimum ellipticity was discussed in \cite{MinimumEllipticity}. A comprehensive review of continuous gravitational wave emission mechanisms from neutron stars can be found in \cite{CWReviewMechanisms}. The detection of a CW, possibly combined with electromagnetic observations of the same source, could yield insight into the structure of neutron stars and into the equation of state of matter under extreme conditions.

Searches for continuous waves are usually split in three different domains: targeted searches look for signals from known pulsars~\cite{S1Paper, S2TDPaper, S3S4TDPaper, Crab, S5TDPaper, Vela, InitialEra, NarrowBand, NarrowBand2, O1KnownPulsars, Polarisations}; directed searches look for signals from known sky locations like the Galactic Center, supernova remnants and low-mass X-ray binaries such as Sco-X1~\cite{S2FstatPaper, CasA, GC, ScoX1Initial, S6SNRPaper, orionspur, SNR1987A, ScoX1S6, S6GlobularCluster, ScoX1O1, ScoX1O1CrossCorr}; all-sky searches look for signals from unknown sources~\cite{S2Hough, S4CWAllSky, S4EH, FullS5Semicoherent, FullS5EH, S5Hough, VSR1TDFstat, S6TwoSpect, S6PowerFlux, ref:VSRFH, S6BucketEH, O1CWAllSkyLowFreq, O1EH, O1CWAllSkyFull}. Since all-sky searches need to cover a large parameter space, they are the most computationally expensive. For this reason, the most sensitive coherent search methods (e.g. matched filtering for the full observing run) cannot be used and semi-coherent methods which split the full observation time in shorter chunks need to be used.

The interest in all-sky searches stems from the fact that they inspect regions of parameter space that no other searches look at. Although the targeted searches are more sensitive, they are limited to search for known pulsars which are in the sensitive frequency band of the detectors, while all-sky searches look for neutron stars with no electromagnetic counterpart, which could have different or more extreme properties than the observed pulsars.

In this paper we present the results of an all-sky search of CWs by three different pipelines (\textit{FrequencyHough} \cite{ref:FH1}, \textit{SkyHough} \cite{SkyHoughMethod}, \textit{Time-Domain $\mathcal{F}$-statistic} \cite{Fstat}) using O2 data from the Advanced LIGO detectors. Each pipeline uses different data analysis methods and covers different regions of the parameter space, although there exists some overlap between them. Overall, we search the whole sky for gravitational wave frequencies from 20 to 1922 Hz (this number was chosen in order to avoid the violin modes of the test masses found at higher frequencies) and a first frequency derivative from $-1\times10^{-8}$ to $2\times10^{-9}$ Hz/s (positive frequency derivatives are possible for neutron stars which are spun up by accretion from a companion). No detection has been made, and upper limits on the gravitational wave amplitude are presented.

The outline of the paper is the following: in Section II, we summarize the second observing run and give some details about the data that is used; in Section III, we describe the model of the signal that we want to detect; in Section IV, we present the different pipelines which are used; in Section V, we describe the results obtained by each pipeline; in Section VI, we give some final remarks.

\section{LIGO O2 observing run}

The LIGO second observing run (called O2) started on November 30 2016 and finished on August 25 2017. During this time, three different gravitational-wave detectors of second generation were active and producing data: Advanced LIGO \cite{AdvancedLIGO}, consisting of two detectors with 4-km arm lengths situated in Hanford, Washington (H1) and Livingston, Louisiana (L1), and Advanced Virgo \cite{AdvancedVirgo}, a 3-km detector located in Cascina, Pisa. Advanced Virgo first joined the run at the beginning of August 2017, with less sensitivity than the LIGO detectors, so we have not considered its data for the search described in this paper. 

\begin{figure*}[htbp]
\begin{center}
\includegraphics[width=1.0\columnwidth]{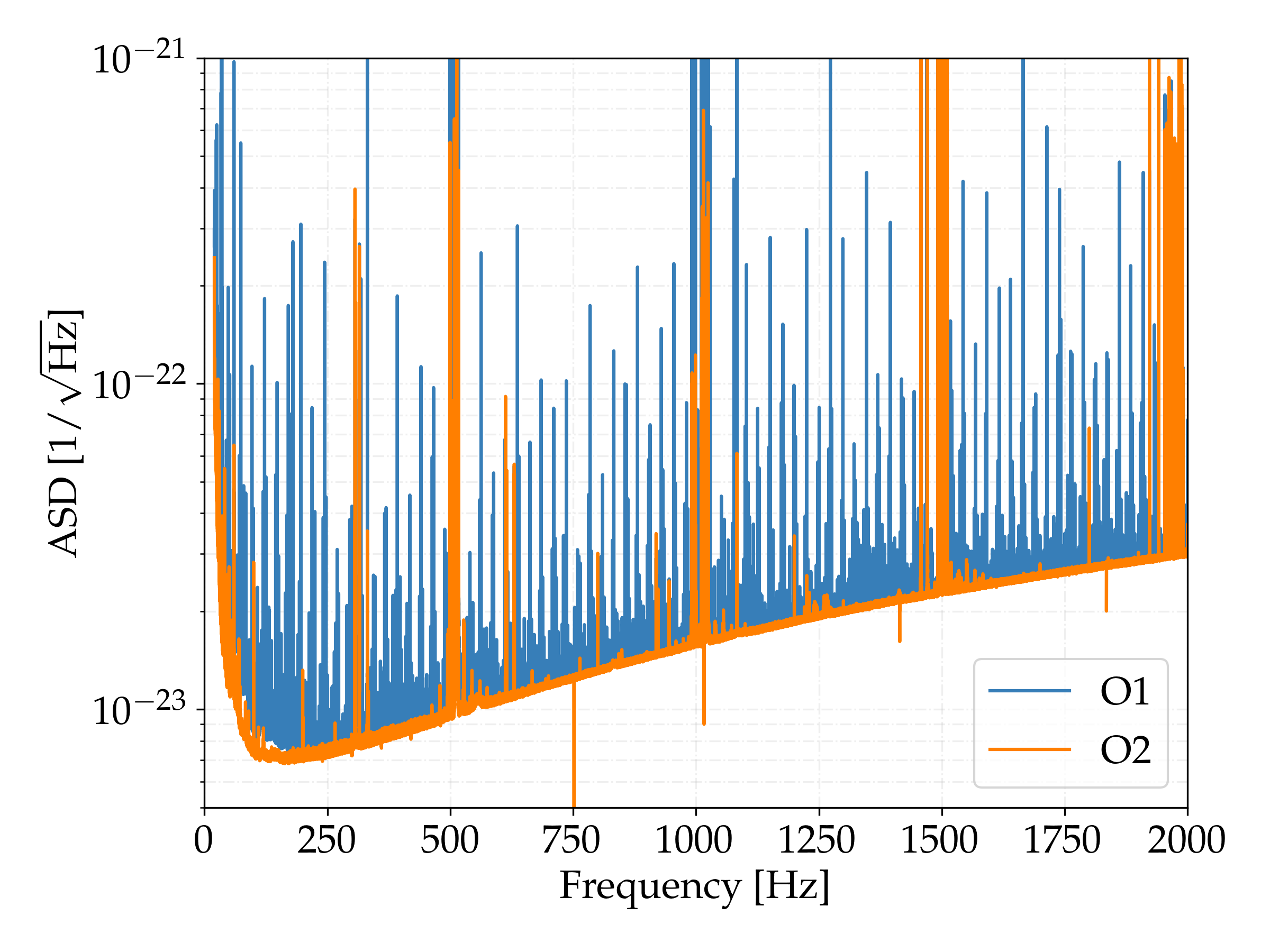}
\includegraphics[width=1.0\columnwidth]{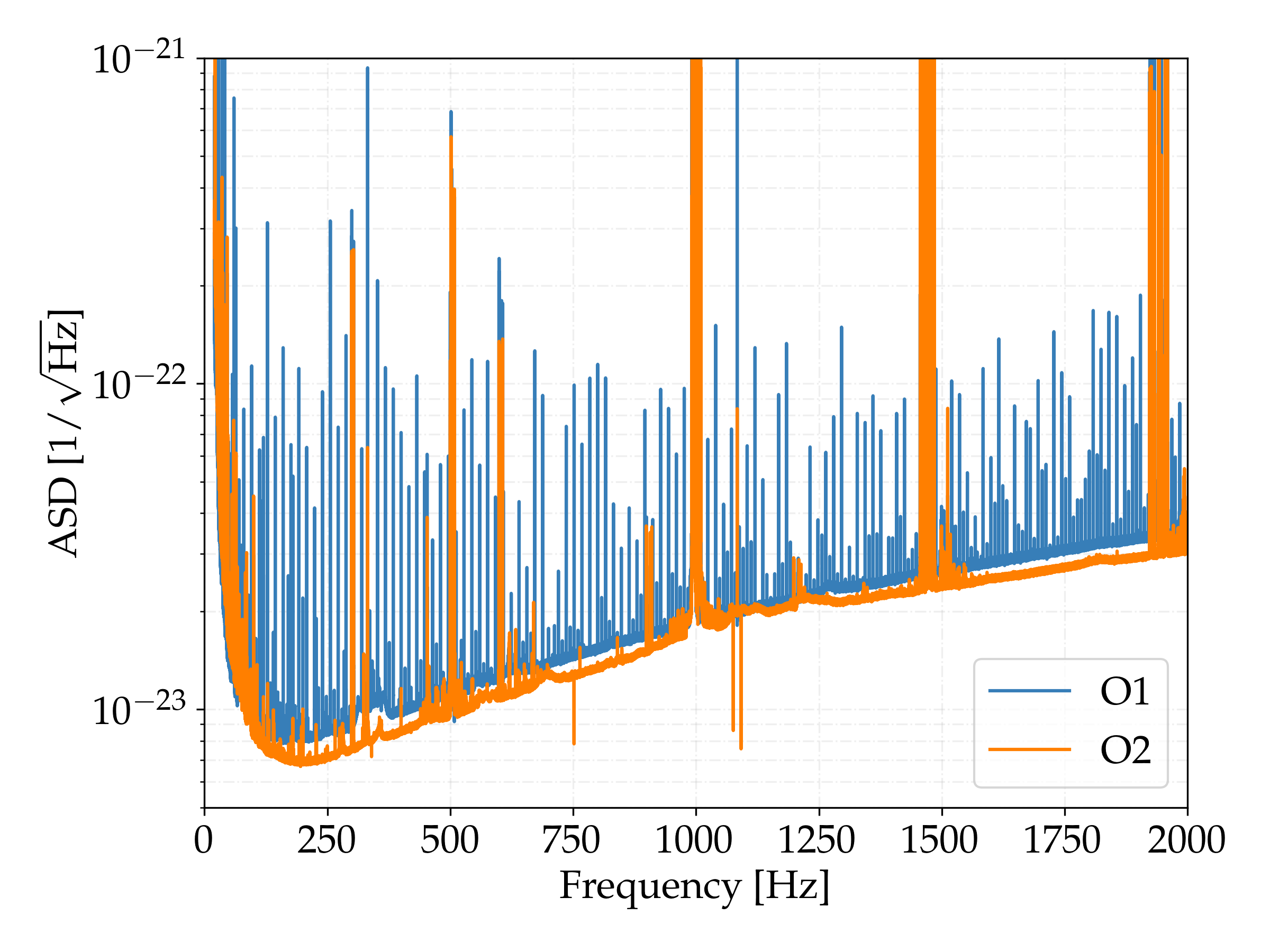}
\caption{Amplitude spectral density (ASD) $\sqrt{S_n}$ plots for the L1 (left panel) and H1 (right panel) detectors during O1 (blue trace) and O2 (orange trace). The ASD is obtained by averaging over FFTs of 1800 s obtained for the entire run.}
\label{fig:ASD}
\end{center}
\end{figure*}

A representative noise curve from O2 for each LIGO detector and a comparison to O1 is shown in Fig. \ref{fig:ASD}. We can observe an improvement of the amplitude spectral density, and we can also observe that the spectra features a number (greatly reduced as compared to O1) of narrow lines and combs affecting several frequency bands, which contaminate the data and complicate the analysis often raising outliers which look like the searched CW signals \cite{LinesCombs}. A cleaning procedure was applied to H1 data during post-processing in order to remove jitter noise and some noise lines (more details are given in \cite{Cleaning}). All of the searches of this paper used this cleaned dataset. The calibration of this dataset and its uncertainties on amplitude and phase are described in \cite{Calibration}. These searches don't use all the data from the observing run, since times where the detectors are known to be poorly behaving are removed from the analysis. This means that the effective amount of data used is smaller than the full duration of the run. As in previous observing runs, several artifical signals (called hardware injections) have been physically injected in the detectors in order to test their response and to validate the different pipelines. These hardware injections are described in Sections III, V and in the Appendix.

\section{Signal model}\label{ref:signal}

An asymmetric neutron star spinning around one of its principal axis of inertia emits a CW signal at twice its rotation frequency. This emission is circularly polarized along the rotation axis and linearly polarized in the directions perpendicular to the rotation axis. The gravitational-wave signal in the detector frame is given by \cite{Fstat}:
\begin{align}
    h(t) = h_0[F_+(t)\frac{1+\cos{\iota}}{2} \cos{\phi(t)} + F_{\times}(t) \cos{\iota} \sin{\phi(t)} ],
    \label{eq:h0t}
\end{align}
where $F_+(t)$ and $F_{\times}(t)$ are the antenna patterns of the detectors (which can be found in \cite{Fstat}), $h_0$ is the amplitude of the signal, $\iota$ is the inclination of the neutron star angular momentum vector with respect to the observer's sky plane, and $\phi(t)$ is the phase of the signal. The amplitude of the signal is given by:
\begin{align}
        h_0 = \frac{4\pi^2G}{c^4} \frac{I_{zz} \epsilon f^2}{d},
        \label{eq:h0}
\end{align}
where $d$ is the distance from the detector to the source, $f$ is the gravitational-wave frequency, $\epsilon$ is the ellipticity or asymmetry of the star, given by $(I_{xx}-I_{yy})/I_{zz}$, and $I_{zz}$ is the moment of inertia of the star with respect to the principal axis aligned with the rotation axis. These two last quantities are related to the mass quadrupole moment $Q_{22}$ of the star:
\begin{align}
    \epsilon = \sqrt{\frac{8\pi}{15}} \frac{Q_{22}}{I_{zz}}.
\end{align}

We assume that the phase evolution of the gravitational-wave signal, which is locked to the evolution of the rotational frequency, can be approximated with a Taylor expansion (assumption taken from electromagnetic observations of pulsars) around a fiducial reference time $\tau_r$:
\begin{align}
    \phi(\tau) = \phi_0 + 2\pi [f_0 (\tau-\tau_r) + \frac{\dot{f}}{2!} (\tau-\tau_r)^2 + ...],
    \label{eq:phaseevo}
\end{align}
where $\phi_0$ is an initial phase and $f_0$ and $\dot{f}$ are the frequency and first frequency derivative at the reference time. The relation between the time at the source $\tau$ and the time at the detector $t$ is given by (neglecting relativistic effects like the Einstein and Shapiro delays):
\begin{align}
    \tau (t) = t + \frac{\vec{r}(t)\cdot\hat{n}}{c},
\end{align}
where $r(t)$ is the vector which joins the Solar System Barycenter (SSB) and the detector, and $\hat{n}$ is the vector identifying the star's position in the SSB. From the previous formula the frequency evolution of the signal can be derived as:
\begin{align}
    f(t) = \frac{1}{2\pi}\frac{d\phi}{dt} \simeq f_0 + f_0\frac{\vec{v}(t)\cdot\hat{n}}{c}+ \dot{f} t.
    \label{eq:frequencyevo}
\end{align}
The second term in the right-hand side of this equation describes the frequency modulation due to the Doppler effect produced by Earth's rotation and translation around the SSB. This term, together with the spin-down/up of the source, must be properly taken into account when carrying out the search.

Equations \ref{eq:phaseevo} through \ref{eq:frequencyevo} assume that the neutron star is isolated. In case it is part of a binary system, the frequency evolution is complicated by the binary system orbital motion, which introduces an additional frequency modulation. Such modulation, on a signal of frequency $f$ and neglecting the binary system ellipticity, is given by (see \cite{Leaci2017}):
\begin{align}
    \Delta f_{\text{orb}}\simeq \frac{2\pi}{P}a_pf,
    \label{eq:binary1}
\end{align}
where $P$ is the binary orbital period and $a_p$ is the projected orbital semi-major axis (in light-seconds). By imposing that the orbital frequency modulation is contained into a frequency bin $\delta f=1/T_{FFT}$, where $T_{FFT}$ is the duration of the data chunks which are incoherently combined in the analysis (see Section~\ref{sec:methods}), we find that two of the search pipelines (\textit{FrequencyHough} and \textit{SkyHough}) used in this paper would be fully sensitive to a CW signal from a NS in a binary system if:
\begin{align}
    a_p \ll 0.076\left(\frac{P}{1~\text{day}}\right)\left(\frac{f}{100~\text{Hz}}\right)^{-1}\left(\frac{T_{FFT}}{1800~\text{s}}\right)^{-1}~\text{s}.
    \label{eq:binary2}
\end{align}
For larger orbital frequency modulations the pipelines would start to lose signal-to-noise ratio but a detailed study of this issue is outside the scope of the paper. Out of 259 pulsars in binary systems from the ATNF catalogue, only 6 of them have such characteristics, although many undiscovered neutron stars in binary systems could also be part of systems with these properties.

\section{Description of the search methods}\label{sec:methods}

In this Section we introduce and summarize the three different pipelines which have been used in this work.

\subsection{FrequencyHough}\label{sec:fh_method}
The \fh\ pipeline consists of an initial multi-step phase, in which interesting points (i.e. candidates) are selected in the signal parameter space, and of a 
subsequent follow-up 
stage to confirm or reject the candidates.  A complete description of the method and of the fundamental implementation features are given in  \citep{ref:FH0,ref:FH1}. Upper limits are computed with a frequentist approach, by injecting a large number of simulated signals into the data. The pipeline
has been previously used in all-sky searches of Virgo VSR2, VSR4~\citep{ref:VSRFH} and LIGO O1 Science Runs data~\cite{O1CWAllSkyLowFreq}.

\subsubsection{Initial analysis steps }\label{sec:fh_steps}
The starting point of the analysis are calibrated detector data, used to create ``short duration'' Fast Fourier Transform (FFTs) with coherence time depending on the frequency band being considered, according to Table~\ref{tab:fh_fft}. Short time-domain disturbances are removed from the data before constructing the FFTs~\cite{peakmap}.
Next, a time-frequency map, called peakmap, is built by identifying local maxima (called peaks) above a dimensionless threshold  $\theta_{\text{thr}}=1.58$ on the square root of the equalized power\footnote{Defined as the ratio of the squared modulus of the FFT of the data and an auto-regressive estimation of the power spectrum, see \cite{peakmap} for more details.} of the data~\cite{peakmap}. The peakmap is cleaned using a line {\it persistency} veto~\citep{ref:FH1}, based on the projection of the peakmap onto the frequency axis and on the removal of the frequency bins in which the projection is higher than a given threshold. 

\begin{table}[tbp]
\begin{center}
\begin{tabular}{ccccc}\hline
Band [Hz] & $T_{\text{FFT}}$ [s] &  $\delta f$ [Hz] & $\delta \dot{f}$ [Hz/s]\\
\hline \hline
  $10$--$128$ & 8192 & $1.22\times 10^{-4}$ & $5.26\times 10^{-12}$ \\
  $128$--$512$ & 4096 &  $2.44\times 10^{-4}$ & $1.05\times 10^{-11}$ \\
  $512$--$1024$ & 2048 &  $4.88\times 10^{-4}$ & $2.10\times 10^{-11}$ \\
  $1024$--$2048$ & 1024 &  $9.76\times 10^{-4}$ & $4.20\times 10^{-11}$ \\
 \hline
\end{tabular}
\caption[FH FFT]{Properties of the FFTs used in the \fh\ pipeline. The time duration $T_{\text{FFT}}$ refers to the length in seconds of the data chunks on which the FFT is computed. The frequency bin width is the inverse of the time duration, while the spin-down/up bin width is computed as $\delta \dot{f}={\delta
f}/{T_{\textrm{obs}}}$, where $T_{\textrm{obs}}$ is the total run duration. In the analysis described in this paper only the first three bands have been considered, the last one will be analyzed in a future work. The spin-down/up range covered by the analysis is ($+2\times 10^{-9}\textrm{ Hz}/\textrm{s}$ , $- 10^{-8}\textrm{ Hz}/\textrm{s}$) up to 512 Hz and ($+2\times 10^{-9}\textrm{ Hz}/\textrm{s}$ , $- 2\times 10^{-9}\textrm{ Hz}/\textrm{s}$) from 512 Hz up to 1024 Hz.}
\label{tab:fh_fft}
\end{center}
\end{table}

After defining a grid in the sky, with bin size depending on the frequency and sky location as detailed in \cite{ref:FH1},  the time-frequency peaks are properly shifted, for each sky position, to compensate the Doppler effect due to the detector motion, see Eq.(\ref{eq:frequencyevo}). They are then processed by the
\fh\ algorithm~\citep{ref:FH0,ref:FH1}, which transforms each peak to the frequency and spin-down/up plane of the source. The frequency and spin-down/up bins (which we will refer to as {\it coarse} bins in the following) depend on the frequency band, as indicated in Table ~\ref{tab:fh_fft}, and are defined, respectively, as $\delta f =\frac{1}{T_{\text{FFT}}}$ and  $\delta \dot{f}={\delta
f}/{T_{\textrm{obs}}}$, where $T_{\textrm{obs}}=268.37$ days is the total run duration. In practice, as the transformation from the peakmap to the Hough plane is not computationally bounded by the width of the frequency bin, we have increased the nominal frequency resolution by a factor of 10 \cite{ ref:FH1}. The algorithm, moreover, properly weights any noise non-stationarity and the time-varying detector response~\citep{ref:Hough_adap}.  

The \fh\ transform is computationally very demanding, so the analysis is split into tens of thousands of independent jobs, each of which computes a \fh\ transform covering a
small portion of the parameter space. The output of a \fh\ transform is a 2-D histogram in the frequency/spin-down plane of the source.
Candidates for each sky location are selected by dividing each 1-Hz band of the corresponding \fh\ histogram into 20 intervals and taking, for each interval, the one or (in most cases) two candidates with the highest histogram number count. This allows us to avoid blinding by large disturbances in the data, as described in \citep{ref:FH1}. All the steps described so far are applied separately to  the data of each detector involved in the analysis.

Following the same procedure used in \cite{O1CWAllSkyLowFreq}, candidates from each detector are clustered and then coincident candidates among the clusters of the two detectors are found using a distance metric built in the four-dimensional
parameter space of position $(\lambda,~\beta)$ (in ecliptic coordinates), frequency $f$ and spin-down/up $\dot{f}$, defined as
\begin{equation}
d_{\rm FH}=\sqrt{\left(\frac{\Delta f}{\delta f}\right)^2+\left(\frac{\Delta \dot{f}}{\delta \dot{f}}\right)^2+\left(\frac{\Delta \lambda}{\delta \lambda}\right)^2+\left(\frac{\Delta 
\beta}{\delta \beta}\right)^2},
\label{eq:dfh}
\end{equation}
where $\Delta f$, $\Delta \dot{f}$, $\Delta \lambda$, and $\Delta \beta$ are the differences, for each parameter, among pairs of candidates of the two detectors, and $\delta f$, $\delta 
\dot{f}$, $\delta \lambda$, and $\delta \beta$ are the corresponding bin widths. Pairs of candidates with distance  $d_{\rm FH}<3$ are considered coincident. This value was chosen based on a study with software simulated signals and allows, on one hand, to reduce the false alarm probability and, on the other, to be sufficiently robust with respect to the fact that a signal can be found with slightly different parameters in the two detectors. Coincident candidates are subject to a ranking procedure, based on the value of a statistic built using the distance and the \fh\ histogram weighted number count of the coincident candidates, as described 
in \citep{ref:FH1}. In this analysis, after the ranking the eight candidates in each 0.1-Hz band with the highest values of the statistic have been selected.
\subsubsection{Candidate follow-up}
\label{sec:fh_followup}
Candidates passing the ranking selection are followed-up in order to confirm them as potential CW signals or to discard them, if due to noise fluctuations or detector disturbances. 
The follow-up consists of several steps, as described in~\cite{ref:VSRFH}. An important implementation novelty we have introduced in the O2 analysis is the use of the {\it Band Sampled Data} framework (BSD) \cite{BSD}, which allows a flexible and computationally efficient management of the data.   
For each of the $N$ candidates selected by the ranking procedure, a fully coherent search is done using down-sampled data from
both detectors, covering a band of the order of 0.2 Hz around the candidate frequency.  The coherent search is done by applying a Doppler and spin-down/up correction based on the parameters of the candidate.  Although the coherent search corrects exactly for the Doppler and spin-down/up effect at a particular point in the parameter space, corresponding to the candidate, the correction is extended by linear interpolation to the neighbors of the candidate itself. In practice, this means that from the corrected and down-sampled time series, a new set of FFTs is built, with a longer duration (by a varying factor $\mathcal{E}$, depending on the frequency band, see Table \ref{tab:fh_FUlength}), as well as the corresponding peakmap. Peaks are selected using a threshold $\theta_{\text{thr}}$=1.87, bigger than the initial one (see Section \ref{sec:fh_method}).  As explained in \cite{ref:VSRFH} this is a conservative choice which provides a sensitivity gain and, at the same time, reduces the computational cost of the follow-up.
\begin{table}[tbp]
\begin{center}
\begin{tabular}{ccccccc}
\hline
Band & FFT duration & $\mathcal{E}$ &  $\delta f$  & $\delta \dot{f}$  & $\mathcal{G}$ \\
\multicolumn{1}{c}{[Hz]}   & \multicolumn{1}{c}{[s]}  &              & [Hz]         & [Hz/s]            & \\
\hline \hline
  $10 $--$ 128$ & 24600 & 3 & $4.07\times 10^{-5}$ & $1.75\times 10^{-12}$ & 1.39\\
  $128 $--$ 512$ & 24576 &  6 & $4.07\times 10^{-5}$ & $1.75\times 10^{-12}$ & 1.65 \\
  $512 $--$ 1024$ & 8192 &  4 & $1.22\times 10^{-4}$ & $5.26\times 10^{-12}$ & 1.49\\
\hline
\end{tabular}
\caption[FH FFT]{Properties of the FFTs used in the \fh\ follow-up step. The second column is the increased FFT duration, the third is the enhancement factor $\mathcal{E}$, with respect to the original duration. The fourth and fifth columns show, respectively, the new frequency and spin-down/up bins, while the sixth is the estimated sensitivity gain $\mathcal{G}$. The new durations have been chosen in such a way to avoid the effect of the sidereal modulation, which produces a spread of the signal power in frequency sidebands \cite{SID}. Actually, for the third band we have used a shorter duration due to computer memory constraints. The last band, from 1024 Hz to 2048 Hz, has not been considered in this work.}
\label{tab:fh_FUlength}
\end{center}
\end{table}
As a result of the FFT length increase and of the new threshold for the selection of the peaks, by using Eq. (67) of \cite{ref:FH1}, which is valid under the assumption of Gaussian noise, we estimate a sensitivity gain $\mathcal{G}$ for the detectable $h_0$ in the follow-up, shown in the last column of Table \ref{tab:fh_FUlength}.
A small area, centered around the candidate position, is considered for the follow-up. It covers $\pm 3$ coarse bins, which amounts to $7$ bins in each dimension, and thus $49$ coarse sky positions for each candidate. A refined sky grid is built over this area, with an actual number of points which depends on the frequency band and is, on the average, given by  $49\mathcal{E}^2$.
 For each sky position in this refined grid, we evaluate the residual Doppler modulation (with respect to the center of the grid), which is corrected in the peakmap by properly shifting each peak.
 The \fh\ of the resulting ensemble of corrected peakmaps is computed over a frequency and spin-down/up ranges covering $\pm 3$ coarse bins around the candidate 
values. As before, an over-resolution factor of 10 is used for the frequency.
 The absolute maximum identified over all the Hough maps provides the refined parameters of the candidate we are considering. Next, to each pair of candidates from the two detectors we apply a series of vetoes, as detailed in the following.
 
 First, we remove the candidates whose median value of the frequency (computed over the full observing time), after the removal of the Doppler and spin-down/up correction, overlaps a known noise line frequency band (i.e. a line due to a detector disturbance, whose instrumental origin has been understood).
 Second, for each detector a new peakmap is computed using the data coherently corrected with the refined parameters of the candidate and then projected on the frequency axis. We take the maximum $A_p$ of this projection in a range of $\pm 2$ coarse bins around the candidate frequency. We divide the rest of the 0.2 Hz band (which we consider the "off-source" region) into $\sim$250 intervals of the same width, take the maximum of the peakmap projection in each of these intervals and sort in decreasing order all these maxima. We tag the candidate as "interesting" and keep it if it ranks first or second in this list for both detectors.
 Surviving candidates are then subject to a consistency test based on their Critical Ratio, defined as $\textrm{CR}={\left(A_p-\mu_p\right)}/{\sigma_p}$, where $\mu_p$ and $\sigma_p$ are the mean and standard deviation of the peakmap projection on the off-source region. Pairs of coincident candidates are removed if their CRs, properly weighted by the detector noise level at the candidate frequency, differ by more than a factor of five. 
Further details on O2 outlier selection and properties are given in Section \ref{sec:fh_results} .
 
\subsubsection{Upper limit computation}
\label{sec:fh_upperlimits}
Upper limits are computed in each 1-Hz band between 20~Hz and 1000~Hz by injecting software simulated signals, with the same procedure used in~\cite{ref:VSRFH}. For each 1~Hz band 
20 sets of 100 signals each are generated, with fixed amplitude within each set and random parameters (sky location, frequency, spin-down/up, and polarization parameters). These are 
generated in time domain and then added to the data of both detectors in the frequency domain. For each injected signal in a set of 100, an analysis is done using the \fh\ 
pipeline over a frequency band of 0.1 Hz around the injection frequency, the full spin-down/up range used in the real analysis, and nine sky points around the injection position~\cite{ref:VSRFH}. Candidates are selected as in the real analysis, except that no clustering is applied, as it would have been affected by the presence of too many signals. We note, however, that clustering is used in the analysis only to reduce the computational cost and does not affect the subsequent steps. After coincidences and ranking, candidates also coincident with the injected signal parameters, within the follow-up volume discussed in Section \ref{sec:fh_followup}, are selected. Those having a critical ratio larger than the largest critical ratio found in the real analysis in the same 1 Hz are counted as {\it detections}. 
For each 1 Hz band, we build the detection efficiency curve, defined as the fraction of detected signals as a function of their amplitude.
The upper limit is given by the signal amplitude such that the detection efficiency is 95\% . In practice, a fit is used in order to interpolate the detection efficiency curve, as described in \citep{O1CWAllSkyLowFreq}.

\subsection{SkyHough}
The \textit{SkyHough} method has been used in other searches using data from the Initial LIGO S2, S4 and S5 and Advanced LIGO O1 observing runs \cite{S2Hough,S4CWAllSky,S5Hough,O1CWAllSkyLowFreq, O1CWAllSkyFull}. Its main description is given in \cite{SkyHoughMethod}. Here we summarize its main characteristics and the new features that have been implemented in this search. The code for the main part of the search is called \textit{lalapps\_DriveHoughMulti} and is part of the publicly available LALSuite package \cite{LALSuite}.

\subsubsection{Initial analysis steps}\label{sec:sh_steps}

This pipeline uses Short Fourier Transforms (SFTs) of the time-domain $h(t)$ as its input data, with a coherent duration of each chunk varying as a function of the searched frequency (as shown in Table \ref{tab:SkyHoughIn2}). It creates peak-grams, which are spectrograms with the normalized power substituted by 1s (if the power is above a certain threshold $\rho_t = 1.6$) and 0s, where the normalized power in a frequency bin is defined as:
\begin{align}
    \rho_k = \frac{|\tilde{x}_k^2|}{\langle n_k \rangle^2}
    \label{HoughPower}
\end{align}
where $\langle n_k \rangle^2$ is estimated with a running median of 101 frequency bins.

\begin{table}[tbp]
\begin{center}
\begin{tabular}{ c c c }
\hline
Frequency [Hz] & $T_c$ [s] & $N_{\text{SFT}}$ \\ 
\hline \hline
$[50, 300)$                & 3600             & 2544 (4755)      \\ 
$[300, 550)$               & 2700             & 3460 (6568)      \\ 
$[550, 1300)$              & 1800             & 5283 (10195)     \\ 
$[1300, 1500)$             & 900              & 10801 (21200)    \\ 
\hline
\end{tabular}
\caption{Coherent times and number of SFTs for each frequency range searched by the \textit{SkyHough} pipeline. The last column shows the number of SFTs per dataset, and the numbers in parenthesis the SFTs used at the second step of the search.}
\label{tab:SkyHoughIn2}
\end{center}
\end{table} 

We use the Hough transform to track the time-frequency evolution of the signal including the Doppler modulation of the signal at the detector. In the first stage the pipeline employs a look-up Table (LUT) approach, taking into account that at a given time the same Doppler modulation is produced by an annulus of sky positions (given by $\Delta \theta$), which correspond to the width of a frequency bin $\Delta f$:
\begin{align}
    \cos{\Delta \theta} = \frac{c}{v(t)} \frac{f(t) - \hat{f} (t)}{\hat{f} (t)} = \frac{c}{v(t)} \frac{\Delta f}{\hat{f} (t)} ,
    \label{eq:SkyHoughDoppler}
\end{align}
where $f(t)$ is the observed frequency at the detector and $\hat{f} = f_0 + f_1t$ is the searched frequency. The algorithm tracks the sky positions which produce observed frequencies with powers above the threshold. It then stacks these sky positions by following the evolution of the source frequency given by the spin-down/up term at different timestamps and produces a final histogram. The LUT approach reuses the same Doppler modulation pattern for different search frequencies (more details in \cite{SkyHoughMethod}), which produces computational savings in exchange for not following the exact frequency-time pattern.

For each template (described by $f_0,\dot{f},\alpha,\delta$) being searched, a detection statistic called number count significance is calculated:
\begin{align}
    s_n = \frac{n- \langle n \rangle}{\sigma_n}
\end{align}
where $\langle n \rangle$ and $\sigma_n$ are the expected mean and standard deviation of the Hough number count $n$ when only noise is present. The number count $n$ is the weighted sum of 1s and 0s, where the weights (which are proportional to the antenna pattern functions and inversely proportional to the power spectral density) were derived in \cite{SkyHoughWeights}. 

The parameter space is separated in 0.1 Hz frequency bands and in small sky-patches. A toplist is calculated for each of these regions, which has the top templates ordered by the number count significance. For the top templates, a second step is performed where instead of using the look-up Table approach the exact frequency path is tracked. At this second step the power significance is also calculated, which is defined as: 
\begin{align}
    s_P = \frac{P - \langle P \rangle}{\sigma_P}
\end{align}
where instead of summing weighted 1s and 0s the weighted normalized powers given by equation \eqref{HoughPower} are summed. This detection statistic improves the sensitivity of \textit{SkyHough} with a very small increase of computational cost. The best 5000 templates per sky-patch and 0.1 Hz band are passed to the second step, and only the best 1000 candidates per sky-patch and 0.1 Hz band are used for the post-processing. Furthermore, at the second step more SFTs are used than in the first step. This is achieved by sliding the initial times of each SFT that was used at the first step, obtaining more SFTs (approximately twice the previous amount), all of them of $T_c$ contiguous seconds.

The grid resolution was obtained in \cite{SkyHoughMethod} and it is given by:
\begin{align}
    \delta f &= \frac{1}{T_c} \\
    \delta \dot{f} &= \frac{1}{T_c T_{\text{obs}}} \\
    \delta \theta &= \frac{c}{v T_c f P_F}
    \label{eq:SkyHoughRes}
\end{align}
where $P_F$ is a parameter which controls the sky resolution grid. In this search we have set $P_F=2$ for all frequencies.

\subsubsection{Post-processing}

The post-processing consists of several steps:
\begin{enumerate}
\item The output of the main SkyHough search is one toplist for each dataset (there are two datasets, each one with data from two detectors, detailed in Table \ref{tab:SkyHoughTimes}) and each region in parameter space. We search for coincidental pairs between these top-lists, by calculating the distance in parameter space and selecting the pairs which are closer than a certain threshold called $d_{\text{co}}$. For the coincidental pairs the centers (average locations in parameter space weighted by significance) are calculated. The parameter space distance is calculated as:
\begin{align}
    d_{SH} = \sqrt{\left(\frac{\Delta f}{\delta f}\right)^2 + \left(\frac{\Delta \dot{f}}{\delta \dot{f}}\right)^2 + \left(\frac{\Delta x}{\delta x}\right)^2 + \left(\frac{\Delta y}{\delta y}\right)^2},
    \label{postdist}
\end{align}
where the numbers in the numerators represent the difference between to templates and the numbers in the denominators represent the parameter resolution (this distance is unitless and is given as a number of bins). The parameters $x$ and $y$ are the Cartesian ecliptic coordinates projected in the ecliptic plane.

\item Search for clusters in the obtained list of centers. This will group different outliers which can be ascribed to a unique physical cause, and will reduce the size of the final toplist. Again, we set a threshold in parameter space distance (called $d_{\text{cl}}$) and we find candidates which are closer than this distance.

\item Finally, we calculate the centers of the clusters. This is done as a weighted (by significance) average, taking into account all the members of the cluster. We order the obtained clusters in each 0.1 Hz by their sum of the power significance of all the members of a cluster, and we select the highest ranked cluster per 0.1 Hz band, if any. This produces the final list of clusters with their parameters ($f_0,\dot{f},\alpha,\delta$) which will be the outliers to be followed-up.

\end{enumerate}

\subsubsection{Follow-up}
\label{SkyHoughFollowup}

We describe a follow-up method based on the $\mathcal{F}$-statistic (described in more detail in subsection \ref{sec:TDFstat_method}) and the GCT metric method \cite{GCTMetric}. This method uses the \textit{lalapps\_HierarchSearchGCT} code, part of the publicly available LALSuite \cite{LALSuite}, and it is similar in spirit to the multi-step follow-up methods described in \cite{E@HFollowup} or \cite{MCMCFollowup}.

The goal is to compare the $\mathcal{F}$-statistic values obtained from software injected signals to the $\mathcal{F}$-statistic values obtained from the outliers. We want to compare the $\mathcal{F}$-statistic obtained at different stages which scale the coherent time. It is expected that for an astrophysical signal the $\mathcal{F}$-statistic value should increase if the coherent time increases.

The resolution in parameter space is given by \cite{GCTMetric}:
\begin{align}
    \delta f &= \frac{\sqrt{12m}}{\pi T_c} \\
    \delta \dot{f} &= \frac{\sqrt{720m}}{\pi T_c^2 \gamma} \\
    \delta \theta &= \frac{\sqrt{m_{\text{sky}}}}{\pi f \tau_e},
    \label{eq:GCT}
\end{align}
where $m$ and $m_{\text{sky}}$ are mismatch parameters, $\gamma$ is a parameter which gives the refinement between the coherent and semi-coherent stages and $\tau_e = 0.021$ s represents the light time travel from the detector to the center of the Earth.

We now enumerate the different steps of the procedure:
\begin{enumerate}
\item Calculate the semi-coherent $\mathcal{F}$-statistic of outliers with $T_c=7200$ s in a cluster box.
\item Add injections to the original data using a sensitivity depth ($\sqrt{S_n}/h_0$) value which returns $\mathcal{F}$-statistic values similar to the values obtained with the outliers in order to compare similar signals. We inject signals in 8 different frequency bands with 200 injections per band, with a sensitivity depth of $~42$ Hz$^{-1/2}$. Then, search in a small region (around 10 bins in each dimension) around the true parameters of the injections with $T_c=7200$ s. Finally, analyze the distances in parameter space from the top candidates to the injections.
\item Repeat the previous step increasing the coherent time to $T_c=72000$ s.
\item Calculate $\mathcal{F}_{72000s}/\mathcal{F}_{7200s}$ for each of the 1600 injections using the top candidates. The threshold will be the minimum value.
\item Calculate the $\mathcal{F}$-statistic values of outliers with $T_c=72000$ s around the top candidate from the first stage. The size of the window to be searched is estimated from the distances found in step 2.
\item Calculate $\mathcal{F}_{72000s}/\mathcal{F}_{7200s}$ using the top candidate for all outliers. Outliers with values higher than the threshold obtained in step 4 go to the next comparison, and the process is repeated from step 2 increasing the coherent time.
\end{enumerate}

Table \ref{tab:SkyHoughFolInj} summarizes the parameters that we have chosen at each different stage. As we will see in the results Section, only two comparisons between three different stages were needed.

\begin{table}[tbp]
\begin{center}
\begin{tabular}{ c c c c }
\hline
Stage index & $T_c$ [s] & $m$ & $m_{\text{sky}}$ \\ 
\hline \hline
I                & 7200             & 0.1 & 0.01    \\ 
II               & 72000             & 0.1 & 0.003    \\ 
III              & 720000             & 0.1 & 0.0005    \\
\hline
\end{tabular}
\caption{Coherent times and mismatch parameters at each different stage of the \textit{SkyHough} follow-up.}
\label{tab:SkyHoughFolInj}
\end{center}
\end{table}

\subsection{Time-Domain {\Fstat}}
\label{sec:TDFstat_method}

The {\td} search method uses the algorithms described
in~\cite{Fstat,AstoneBJPK2010,VSR1TDFstat,PisarskiJ2015} and has been applied to
an all-sky search of VSR1 data~\cite{VSR1TDFstat} and an all-sky search of the
LIGO O1 data~\cite{O1CWAllSkyLowFreq,O1CWAllSkyFull}.  The main tool is the
{\Fstat} \cite{Fstat} by which one can search coherently the data over a reduced
parameter space consisting of signal frequency, its derivatives, and the sky
position of the source. The {\Fstat} eliminates the need for a grid search over remaining parameters (see Eqs. \ref{eq:h0t} and \ref{eq:phaseevo}), in particular, the inclination angle $\iota$ and polarization $\psi$. Once a signal is identified the estimates of those four parameters are obtained from analytic formulae.  

However, a coherent search over the whole LIGO O2 data set is
computationally prohibitive and we need to apply a semi-coherent method, which
consists of dividing the data into shorter time domain segments. The short time
domain data are analyzed coherently with the {\Fstat}. Then the output from the
coherent search from time domain segments is analyzed by a different,
computationally-manageable method. Moreover, to reduce the computer memory
required to do the search, the data are divided into narrow-band segments that
are analyzed separately.  Thus our search method consists primarily of two
parts. The first part is  the coherent search of narrow-band, time-domain
segments. The second part is the search for coincidences among the candidates
obtained from the coherent search.  

The pipeline is described in Section IV
of~\cite{O1CWAllSkyLowFreq} (see also Fig. 13 of~\cite{O1CWAllSkyLowFreq} for the flow chart
of the pipeline). The same pipeline is used for the search of LIGO O2 data
presented here except that a number of parameters of the search are different.
The choice of parameters was motivated by the requirement to make the search
computationally manageable. 

As in our O1 searches, the data are divided into overlapping frequency sub-bands of
0.25~Hz. We analyze three frequency bands: $[20$-$100]$~Hz,  $[100$-$434]$~Hz,
and  $[1518$-$1922]$~Hz.  As a result, the three bands have 332, 1379, and 1669
frequency sub-bands respectively. 
 
The time series is divided into segments, called frames, of 24 sidereal days
long each, 6 days long, and 2 days long respectively for the three bands.
Consequently in each band we have 11, 44, and 134 time frames, respectively. The
O2 data has a number of non-science data segments. The values of these bad data
are set to zero. For this analysis, we choose only segments that have a
fraction of bad data less than 1/2 both in H1 and L1 data.  This requirement
results in eight 24-day-long, twenty six 6-day-long, seventy nine 2-day-long
data segments for each band respectively.  These segments are analyzed
coherently using the {\Fstat} defined by Eq.~(9) of ~\cite{VSR1TDFstat}. We set
a fixed threshold for the {\Fstat} of {$\mathcal{F}_0 = 16$} and record the
parameters of all threshold crossings, together with the corresponding values
of the signal-to-noise ratio~$\rho$,

\begin{equation}
\rho = \sqrt{2(\F - 2)}.
  \label{eq:tdfstat_O1_fstat_snr}
\end{equation}

Parameters of the threshold crossing constitute a candidate signal. 
At this first stage we also veto candidate signals overlapping with the instrumental
lines identified by independent analysis of the detector data. 

For the search we use a four-dimensional grid of templates (parametrized by
frequency, spin-down/up, and two more parameters related to the position of the
source in the sky) constructed in Section IV of~\cite{PisarskiJ2015}. For the low
frequency band $[20$-$434]$~Hz we choose the grid's minimal match
$\textrm{MM}=\sqrt{3}/2$ whereas for the high frequency band $[1518$-$1922]$~Hz we
choose a looser grid with $\textrm{MM} = 1/2$. 
 
In the second stage of the analysis we search for coincidences among the
candidates obtained in the coherent part of the analysis. We use exactly the
same coincidence search algorithm as in the analysis of VSR1 data and described
in detail in Section VIII of~\cite{VSR1TDFstat}. We search for coincidences in
each of the sub-bands analyzed. To estimate the significance of a given
coincidence, we use the formula for the false alarm probability derived in the
Appendix of~\cite{VSR1TDFstat}.  Sufficiently significant coincidences are
called outliers and subjected to further investigation.

The sensitivity of the search is estimated by the same procedure as in O1 data
analysis (\cite{O1CWAllSkyLowFreq}, Section IV). The sensitivity is taken to be the
amplitude $h_0$ of the gravitational wave signal that can be confidently
detected.  We perform the following Monte-Carlo simulations.  For a given
amplitude $h_0$, we randomly select the other seven parameters of the signal:
$f, \dot{f}, \alpha, \delta, \phi_0, \iota$ and $\psi$.  We choose
frequency and spin-down/up parameters uniformly over their range, and source
positions uniformly over the sky.  We choose angles $\phi_0$ and $\psi$
uniformly over the interval $[0, 2\pi]$ and $\cos\iota$ uniformly over the
interval $[-1, 1]$. We add the signal with selected parameters to the O1 data.
Then the data are processed through our pipeline.  First, we perform a coherent
{\Fstat} search of each of the data segments where the signal was added. Then
the coincidence analysis of the candidates is performed. The signal is
considered to be detected, if it is coincident in more than 5 of the 8 time
frames analyzed,  14 out of 26, and  40 out of 79 for the three bands
respectively. We repeat the simulations one hundred times.  The ratio of
numbers of cases in which the signal is detected to the one hundred simulations
performed for a given $h_0$ determines the frequentist sensitivity upper
limits. We determine the sensitivity of the search in each $0.25$Hz frequency
sub-band separately.  The 95\% confidence upper limits for the whole range of
frequencies are given in Fig.~\ref{fig:tdfstat_O2_upper_limits}; they follow
very well the noise curves of the O2 data that were analyzed. The sensitivity
of search decreases with decreasing coherence time we use for the three bands
of our {\Fstat} search. Additionally it decreases in our high-frequency band
because of the looser grid used than in low frequency bands.

\section{Results}

\begin{figure}[tbp]
\includegraphics[width=1.0\columnwidth]{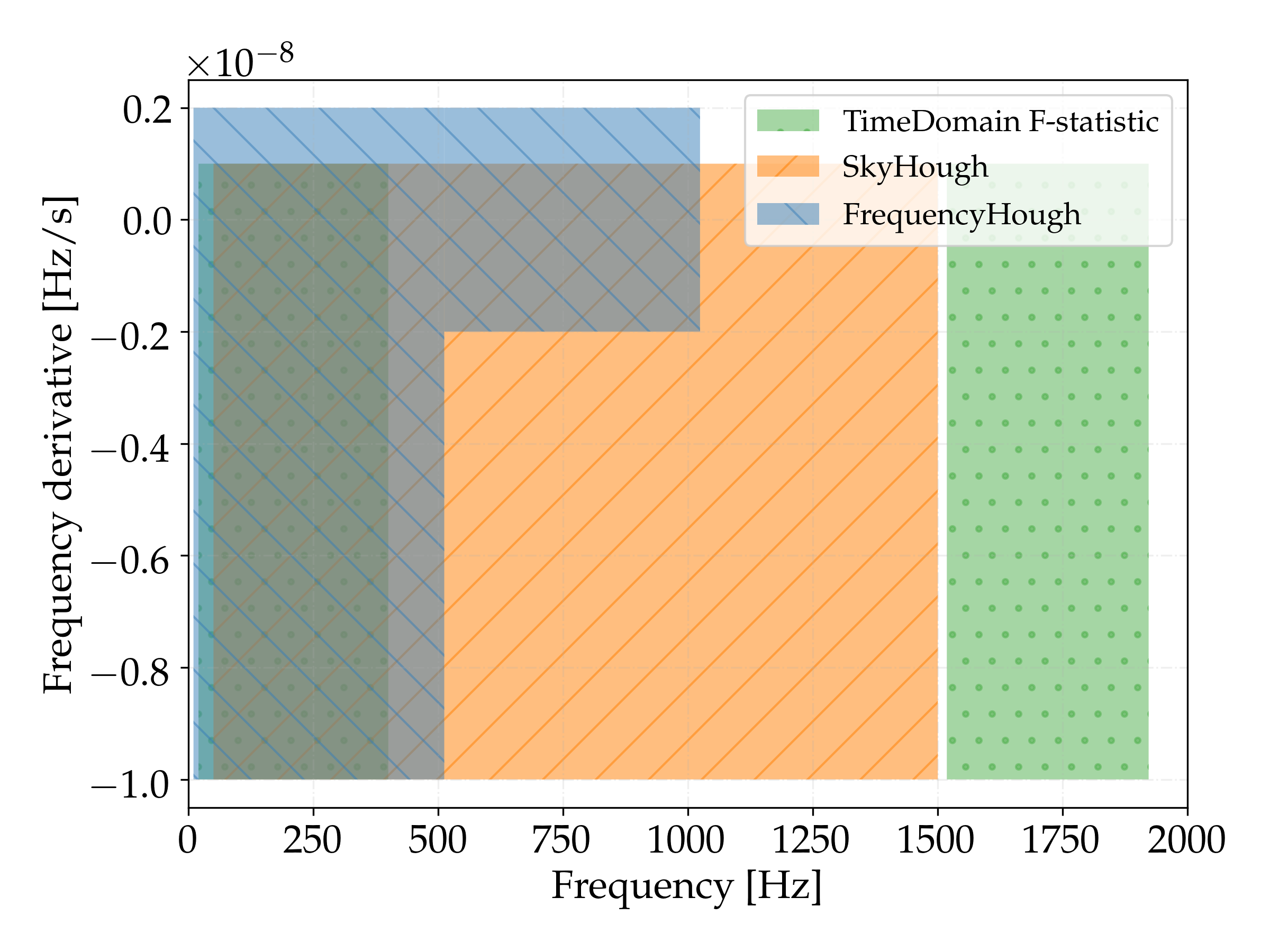}
\caption{Regions in frequency and first frequency derivative covered by each pipeline.}
\label{fig:parameterspacecovered}
\end{figure}

In this Section we detail the results obtained. The region in frequency and first frequency derivative searched by each of the three different pipelines is shown in Fig. \ref{fig:parameterspacecovered}.

\begin{figure*}[tbp]
\includegraphics[width=1.0\linewidth]{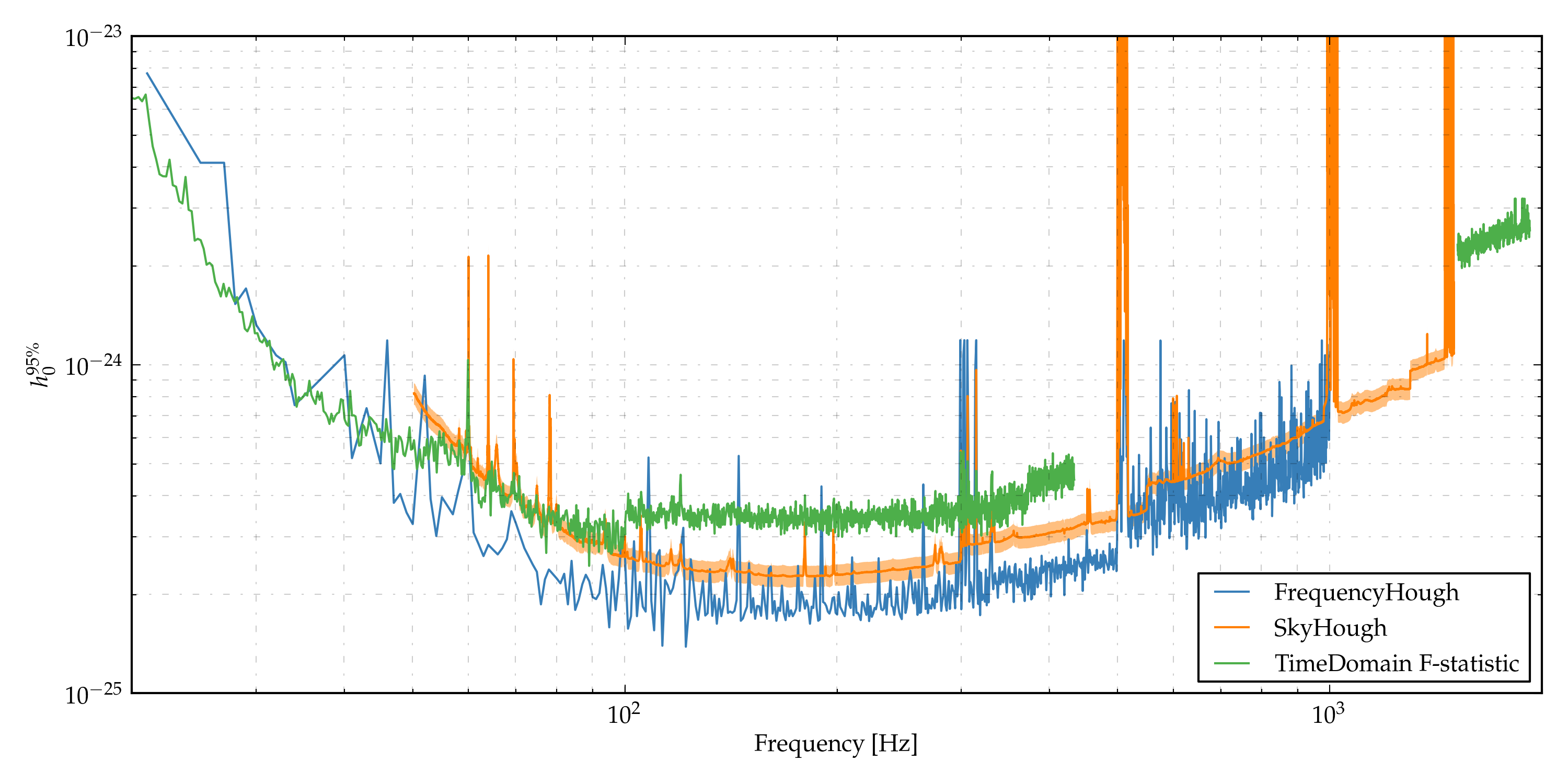}
\caption{Upper limits on the strain amplitude $h_0^{95\%}$ for the three pipelines.} 
\label{fig:globalUL}
\end{figure*}

Although no detections have been made, we give details on the different procedures and outliers which were found, and we also present 95\% confidence level (CL) upper limits on the 
strain $h_0$ given by equation \eqref{eq:h0}, shown in Fig. \ref{fig:globalUL}. The best upper limit is $\simeq 1.7\times10^{-25}$ at around 120 Hz. These results are 
significantly 
better (of a factor of about 1.4) than those obtained on O1 data with the same pipelines~\cite{O1CWAllSkyLowFreq,O1CWAllSkyFull}, thanks to improvements in the pipelines themselves, to the better sensitivity of the detectors and to the longer duration of the observing run. These upper limits do not take into account the calibration uncertainty on amplitude, which over the run was no larger than 5\% and 10\% for H1 and L1 respectively \cite{Calibration}.

Our O2 results are comparable with the upper limits obtained in O1 by the Einstein@Home project~\cite{O1EH} over the range 20-100 Hz. Note, however, that the Einstein@Home search covered a spin-down/up range smaller by almost one order of magnitude. Moreover, while the Einstein@Home search is, in principle, more sensitive due to the use of much longer data segments (compared to the \textit{FrequencyHough} and \textit{SkyHough} pipelines), with 210 hr duration, it is also less robust in case of deviations from the assumed signal model described in Section \ref{ref:signal}. At frequencies higher than 100 Hz, the previous best upper limits were obtained in \cite{Loosely} using O1 data. Our results improve on those upper limits by approximately 17\%. 

\begin{figure*}[htbp]
\includegraphics[width=1.0\columnwidth]{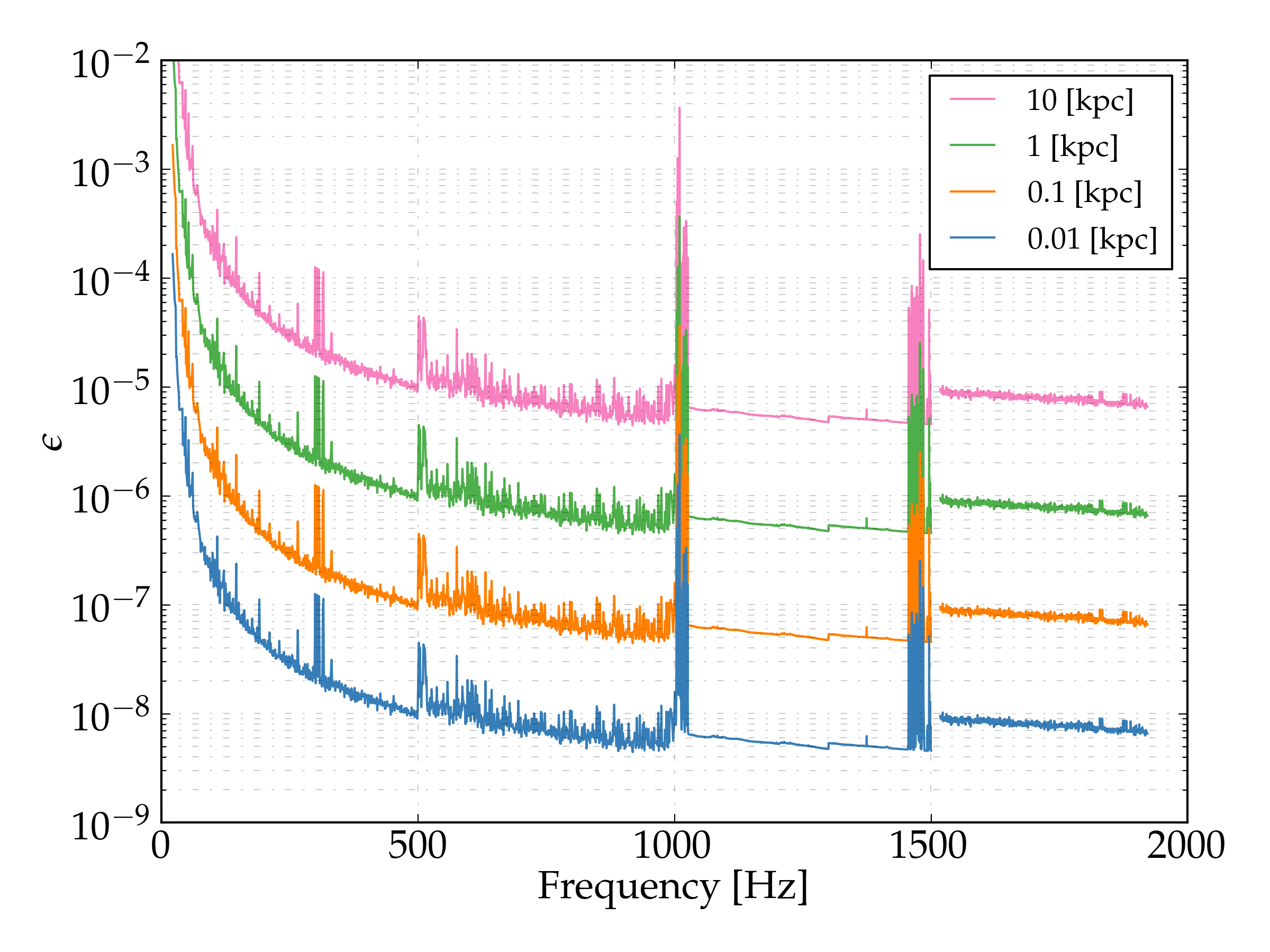}
\includegraphics[width=1.0\columnwidth]{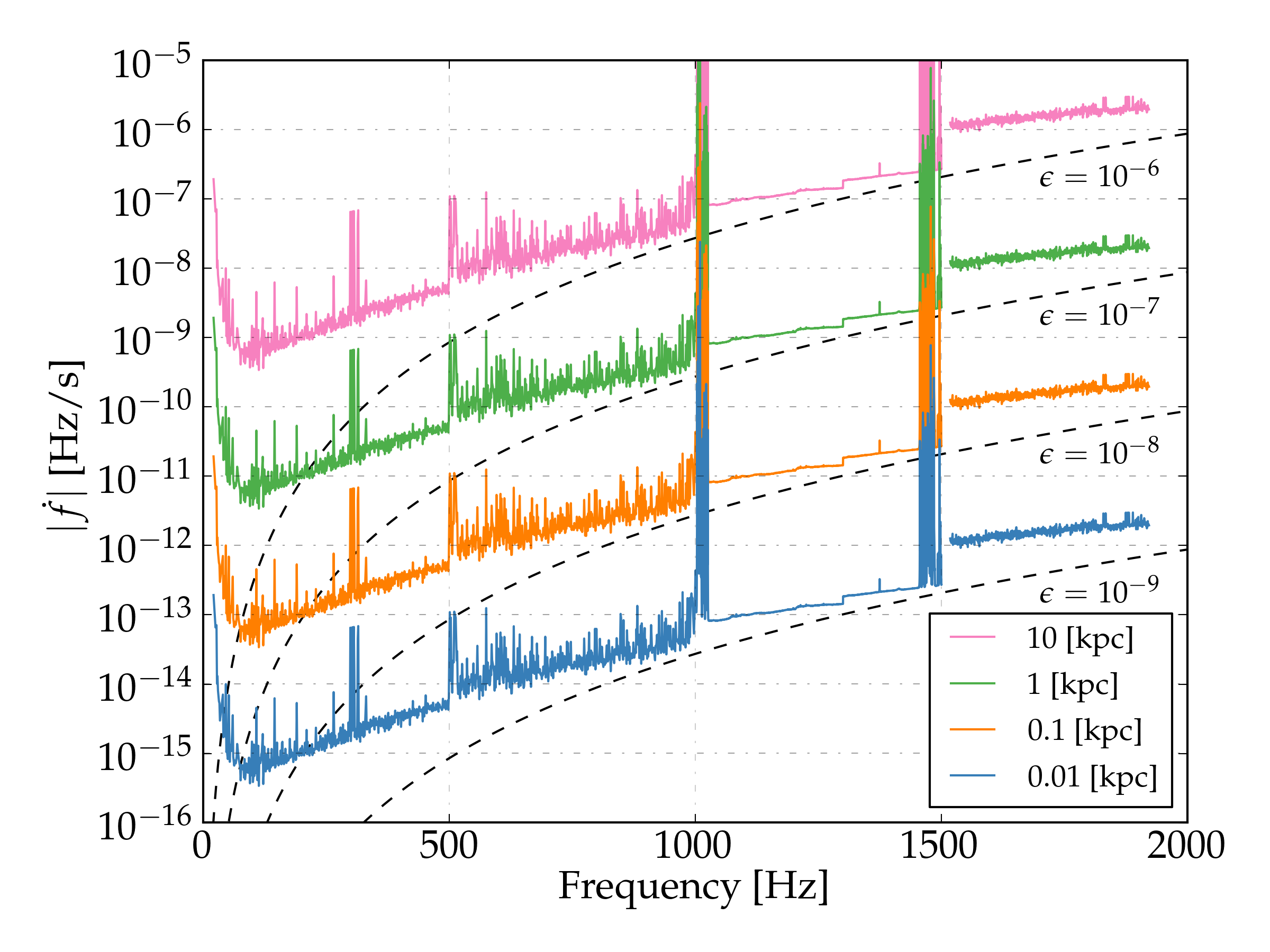}
\caption{The left panel shows the detectable ellipticity given by equation \eqref{eq:eps} as a function of frequency for neutron stars at 10 pc, 100 pc, 1 kpc and 10 kpc for a canonical moment of inertia $I_{zz}=10^{38}$ kg$\cdot$m$^2$. The right panel shows the relation between the absolute value of the first frequency derivative and the frequency of detectable sources as a function of the distance, assuming their spin-down is due solely to the emission of gravitational waves. The different colors correspond to the same distances of the left panel. Black dashed lines are lines of constant source ellipticity, from $\varepsilon =10^{-9}$ (bottom dashed line) to $\varepsilon = 10^{-6}$.}
\label{fig:AstroReach}
\end{figure*}

The $95\%$ CL upper limits on $h_0$ can be converted to upper limits on ellipticity $\epsilon$ by using equation \eqref{eq:h0} with a canonical value for the moment of inertia of $10^{38}$ kg$\cdot$m$^2$ and by using different distances:
\begin{align}
        \epsilon = \frac{c^4}{4\pi^2G} \frac{h_0 d}{I_{zz} f^2}.
        \label{eq:eps}
\end{align}
These results are shown in the left panel of Fig. \ref{fig:AstroReach}. This has been obtained by using the best $h_0$ upper limits between the three pipelines: from 20 to 1000 Hz, the \textit{FrequencyHough} results have been used; from 1000 to 1500 Hz, the results from \textit{SkyHough} have been used; from 1518 to 1922 Hz the results from {\td} have been used. For sources at 1 kpc emitting CWs at 500 Hz, we can constrain the ellipticity at $\simeq 10^{-6}$, while for sources at 10 kpc emitting at the same frequency we can constrain the ellipticity at $10^{-5}$.

A complementary way of interpreting the limits on ellipticity is shown in the right panel of Fig. \ref{fig:AstroReach}. The various set of points give the relation between the absolute value of the signal frequency time derivative (spin-down) and the signal frequency for sources detectable at various distances, assuming their spin-down is only due to the emission of gravitational waves. They have been computed by means of the following relation obtained inverting the equation for the so-called spin-down limit amplitude $h_0^{\text{sd}}$, which is a function of the source distance $d$, frequency $f$ and spin-down $\dot{f}$, see e.g. equation (A7) in \cite{ref:O2targeted}:
\begin{align}\label{eq:ffdoth0}
|\dot{f}|=1.54\times 10^{-10} &\left(\frac{I_{zz}}{10^{38}\,\text{kg} \cdot \text{m}^2}\right)^{-1} \left(\frac{h_0^{\text{sd}}}{10^{-24}}\right) \nonumber \\
&\left(\frac{f}{100\,\text{Hz}}\right) \left(\frac{d}{1\,\text{kpc}}\right)^2~[\mathrm{Hz/s}],
\end{align}
where we have replaced the spin-down limit amplitude with the 95\% upper limits shown in Fig. \ref{fig:globalUL}.
The dashed lines are constant ellipticity curves obtained from equation (A9) of \cite{ref:O2targeted}:
\begin{align}\label{eq:ffdoteps}
|\dot{f}|=1.72\times 10^{-14}&\left(\frac{I_{zz}}{10^{38}\,\text{kg} \cdot \text{m}^2}\right)\left(\frac{f}{100\,\text{Hz}}\right)^{1/2} \nonumber \\
& \left(\frac{\varepsilon}{10^{-6}}\right)^2~[\mathrm{Hz/s}].
\end{align}
For a signal to be detectable, its spin-down/up would need to be equal or above the given traces (notice that, as shown in Fig. \ref{fig:parameterspacecovered}, the maximum absolute spin-down searched is $10^{-8}$ Hz/s, which marks a limit to the signals we are sensitive to). For example, a source emitting a signal with frequency higher than about 500 Hz and ellipticity equal or greater than $10^{-6}$ would be detectable up to a distance of about 1 kpc if its spin-down is, in modulus, larger than $\approx 10^{-10}$ Hz/s.

The three searches carried out by the different pipelines have different computational costs: \textit{FrequencyHough} spent 9 MSU; \textit{SkyHough} spent 2.5 MSU; {\td} spent 24.2 MSU, where 1 MSU hour corresponds to 1 million Intel E5-2670 core-hour to perform a SPECfp computation. We remind the reader that each of these pipelines covered different search bands.

\subsection{FrequencyHough}
\label{sec:fh_results}
In this Section we report the main results of the O2 all-sky search using the \fh\ pipeline.
The spin down range covered by the analysis is ($+2\times 10^{-9}\textrm{ Hz}/\textrm{s}$ $, $ $- 10^{-8}\textrm{ Hz}/\textrm{s}$)  up to 512 Hz and ($+2\times 10^{-9}\textrm{ Hz}/\textrm{s}$ $,$ $- 2\times 10^{-9}\textrm{ Hz}/\textrm{s}$)  from 512 Hz up to 1024 Hz.
    
\begin{figure*}[htbp]
\begin{center}
\includegraphics[width=3.5in]{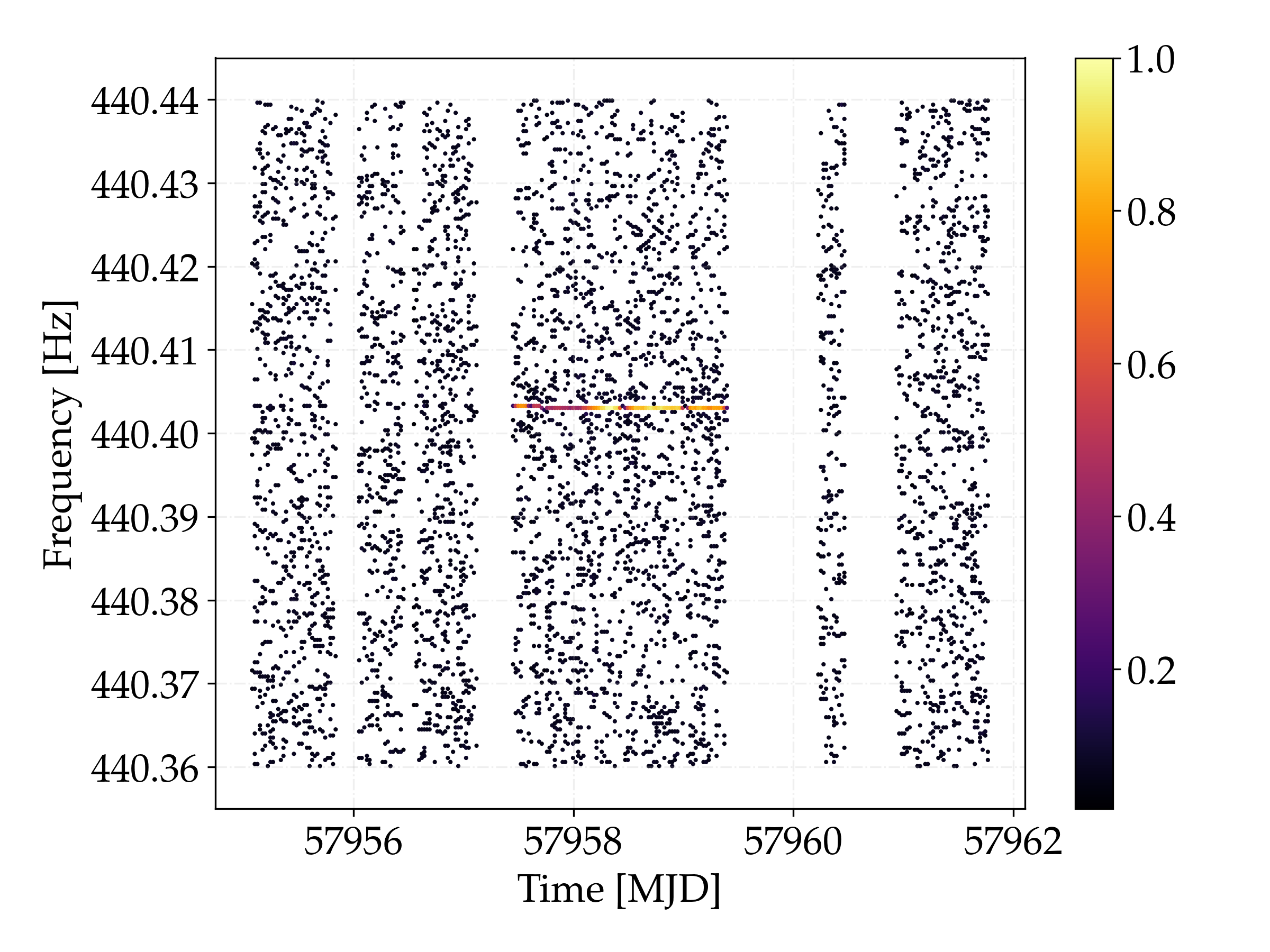}
\includegraphics[width=3.5in]{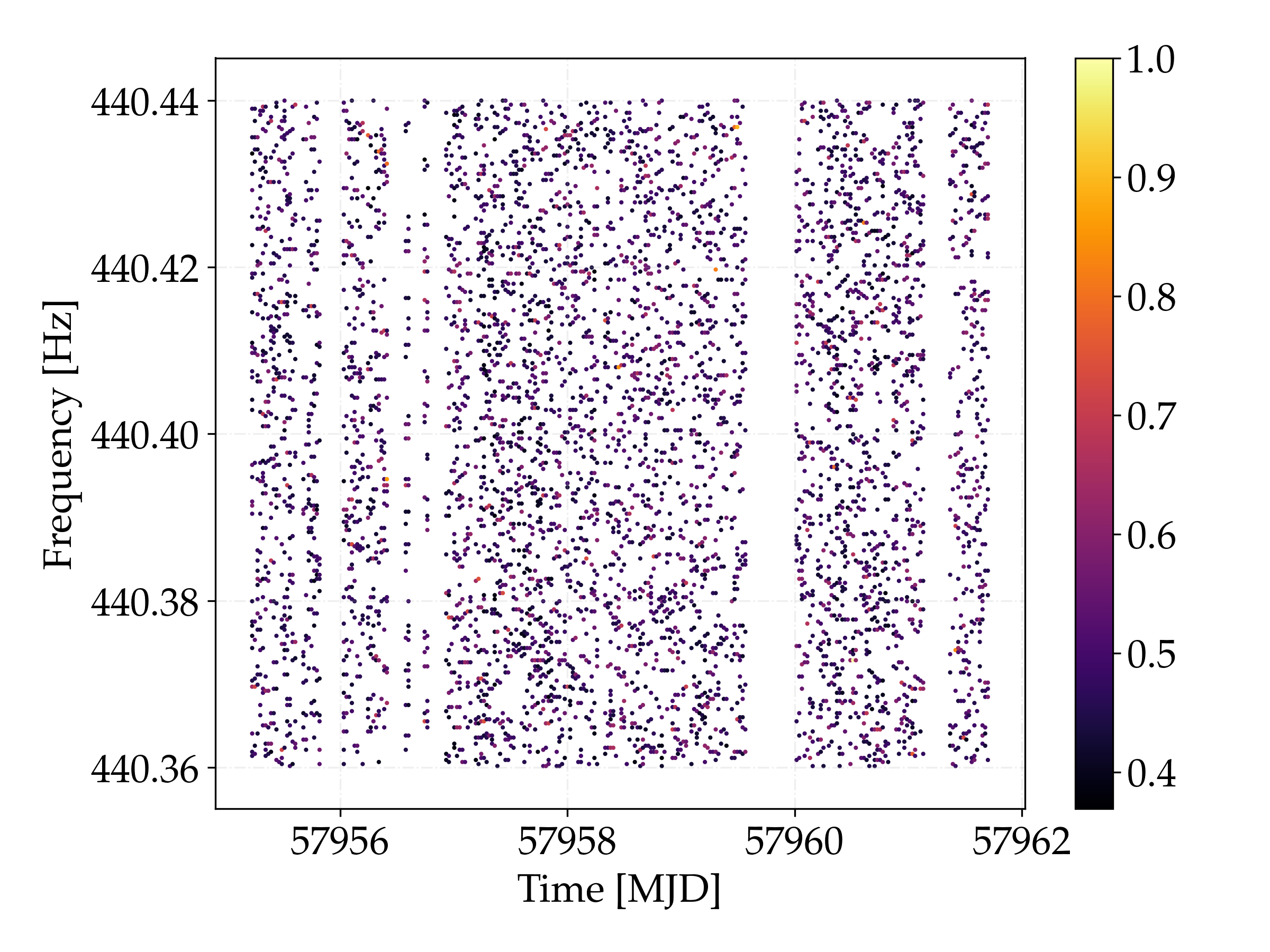}
\caption{\fh\ peakmaps, without Doppler correction, around the outlier at $\sim$440.4 Hz, for Hanford (left) and Livingston (right) data. The presence of a transient line is clearly visible in Hanford. The x-axis indicates time in Modified Julian Date (MJD).}
\label{fig:PM440}
\end{center}
\end{figure*}

The number of initial candidates produced by the \fh\ transform stage was about $5 \times 10^9$ (of which about $7\times 10^7$ belong to the band 20-128 Hz,  about $1.1\times 10^9$  to the band 128-512 Hz and about $3.8 \times 10^9$ to the band 512-1024 Hz, for both Hanford and Livingston detectors.
As the total number of coincident candidates remained too large, $1.09 \times 10^8$, we reduced it with the ranking procedure described in Section 
\ref{sec:fh_method}. The total number of candidates selected after the ranking was 59025.
Each of these candidates was subject to a multi-stage follow-up procedure, described in Section \ref{sec:fh_followup}. The total number of candidates passing the follow-up and all the veto steps
was 154, after removing the candidates due to the hardware injections.  Among these, only 27 were found in coincidence between the two detectors (within a 
distance $d_{\rm FH}<3$ as defined in \eqref{eq:dfh}). From these surviving candidates we selected the outliers less consistent with noise fluctuations. 
In particular, we choose those for which the final peakmap projections have an average (over the two detectors) critical ratio (see Section \ref{sec:fh_followup})  $\textrm{CR}>7.42$. This is the threshold 
corresponding, under the assumption of Gaussian noise, to a false alarm probability of 
$1\%$ after having taken into account the look-elsewhere effect (on the follow-up stage) \cite{O1CWAllSkyLowFreq}.  We found only 1 candidate an average  $\textrm{CR}$ above the threshold.  It was at a frequency of about 440.4 
Hz, and occurred due to a high CR value in the LIGO Hanford detector. For this candidate, we have looked at the starting peakmaps, without 
Doppler correction, around its frequency, which clearly show the presence of a transient line of duration $\approx 2$ days at the candidate frequency in Hanford data, see Fig. 
\ref{fig:PM440}. We have then discarded this candidate as a possible CW signal. 
The remaining 26 sub-threshold candidates, which will be further analyzed in a forthcoming work, are listed in \ref{app:FHST}.
The analysis was run on distributed computational resources accessed through the EGI grid middleware \cite{EGI}.

\begin{figure}[htbp]
\begin{center}
\includegraphics[width=1.0\columnwidth]{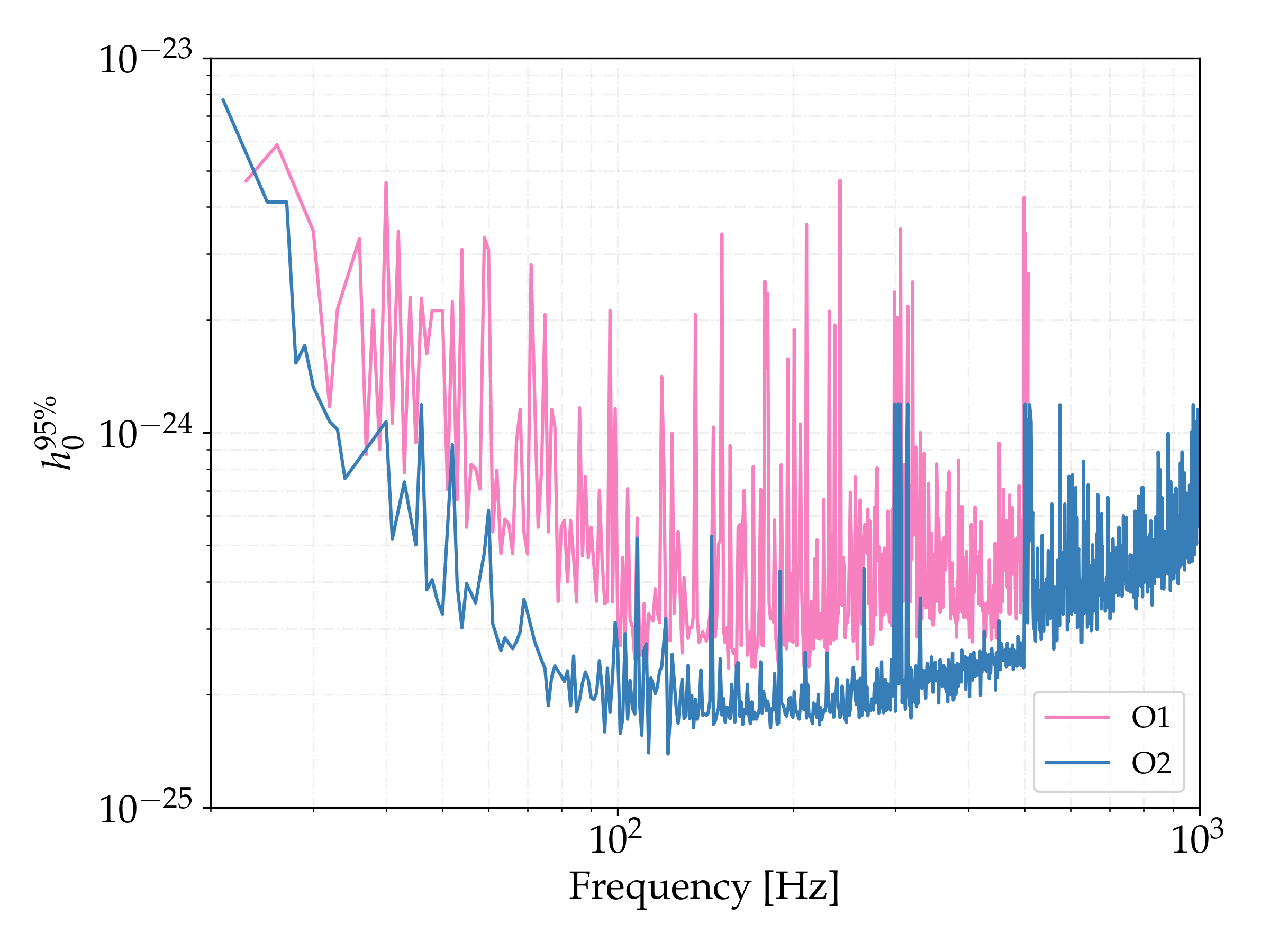}
\caption[Upper limit]{Comparison of O1 and O2 95\% upper limits on the strain amplitude for the \fh\ pipeline. The O2 search covered the range between 20 Hz and 1000 Hz, while the O1 search arrived up to  475 Hz. They have been obtained adding simulated signals to the real data, covering the same parameter space as in the actual search.}
\label{fig:fh_ul}
\end{center}
\end{figure}

As we didn't find any significant candidate, we have computed upper limits.
They have been evaluated in 1-Hz bands, as described in \ref{sec:fh_upperlimits}, and are shown in Fig.~\ref{fig:globalUL}. 
The total amount of frequency bands vetoed by the persistency veto is negligible, as it amounts to less than 0.55 \% and 0.45 \% respectively for LIGO Hanford and Livingston.
There are a few 1 Hz bands where we have not evaluated the upper limit, due to the fact that we don't have candidates or, due to 
disturbances, we have not been able to recover the 95\% of the injections, or more, at any amplitude.
These bands are those with the following initial frequencies: \{22, 23, 24, 26, 35, 36, 37, 39, 42, 51, 56, 65, 71, 73, 79, 120, 763, 995, 996, 998\} Hz.
Comparing upper limits with O1 results \cite{O1CWAllSkyLowFreq}, see Fig.\ref{fig:fh_ul}, we notice an improvement of $\sim 30-40 \%$ at frequencies between $\sim$ 150 and 500 Hz, 
while the gain is significantly bigger, up to a factor of $\sim$ 2, at lower frequencies. This is the first time we have extended the \fh\ analysis above $\sim$ 500 Hz.
The statistical uncertainty on the upper limit is lower than about 5 \%, because of the amplitude step used for the injections, which amounts to $2 \times 10^{-26}$.

As a test of the capabilities of the pipeline to recover signals, we report in Appendix \ref{app:FHHI} the parameters of 
the recovered hardware injections, together with the error with respect to the injected signals. We note that we were able to recover, with very good accuracy, the parameters of all the 12 hardware injections with frequency in the analyzed band.

\subsection{SkyHough}

\textit{SkyHough} has analyzed frequencies from 50 to 1500 Hz and spin-down/up values from $-10^{-8}$ to $10^{-9}$ Hz/s as shown in Fig. \ref{fig:parameterspacecovered}. The four different coherent times that have been used are shown in Table \ref{tab:SkyHoughIn2}. This analysis uses the C02 cleaned dataset \cite{Calibration}, and splits the data from H1 and L1 in two datasets, divided by time as shown in Table \ref{tab:SkyHoughTimes}, where the start and stop times for each dataset are indicated. The main search generates a toplist per dataset per 0.1 Hz band of 10000 candidates with a maximum of 1000 per sky-patch. The number of sky-patches depends on the frequency: to minimize the computational cost of the search, we try to minimize the number of sky-patches for a limited amount of RAM. From 50 to 850 Hz, there are 28 sky-patches; from 850 to 1000 Hz, 31 sky-patches; from 1000 to 1150 Hz, 38 sky-patches; from 1150 to 1250 Hz, 45 sky-patches; from 1300 to 1500 Hz, 28 sky-patches. After applying the post-processing stage previously described (with distance thresholds of $d_{\text{co}}=3$ and $d_{\text{cl}}=\sqrt{14}$), we are left with 4548 0.1 Hz bands (from a total of 14500) having coincidental pairs.

\begin{table}[tbp]
\begin{center}
\begin{tabular}{ c c c }
\hline
\multirow{2}{*}{Dataset 1} & H1 & 1167545839/1174691692 \\ 
 & L1 & 1167546403/1174688389 \\
 \hline
\multirow{2}{*}{Dataset 2} & H1 & 1180982628/1187731792 \\ 
 & L1 & 1179816663/1187731695 \\
\hline
\end{tabular}
\caption{Start/stop times in GPS units of each dataset used by the \textit{SkyHough} pipeline. The observation time parameter used for the spin-down resolution given by equation \eqref{eq:SkyHoughRes} is $T_{obs}=7915032$ s, the maximum span of these datasets.}
\label{tab:SkyHoughTimes}
\end{center}
\end{table}

We apply the \textit{population veto}, used in many past searches, which demands that each dataset contributes to each cluster with at least two different templates. After applying this veto, only 1539 outliers remain. 

\begin{figure*}[htbp]
\begin{center}
\includegraphics[width=1.0\columnwidth]{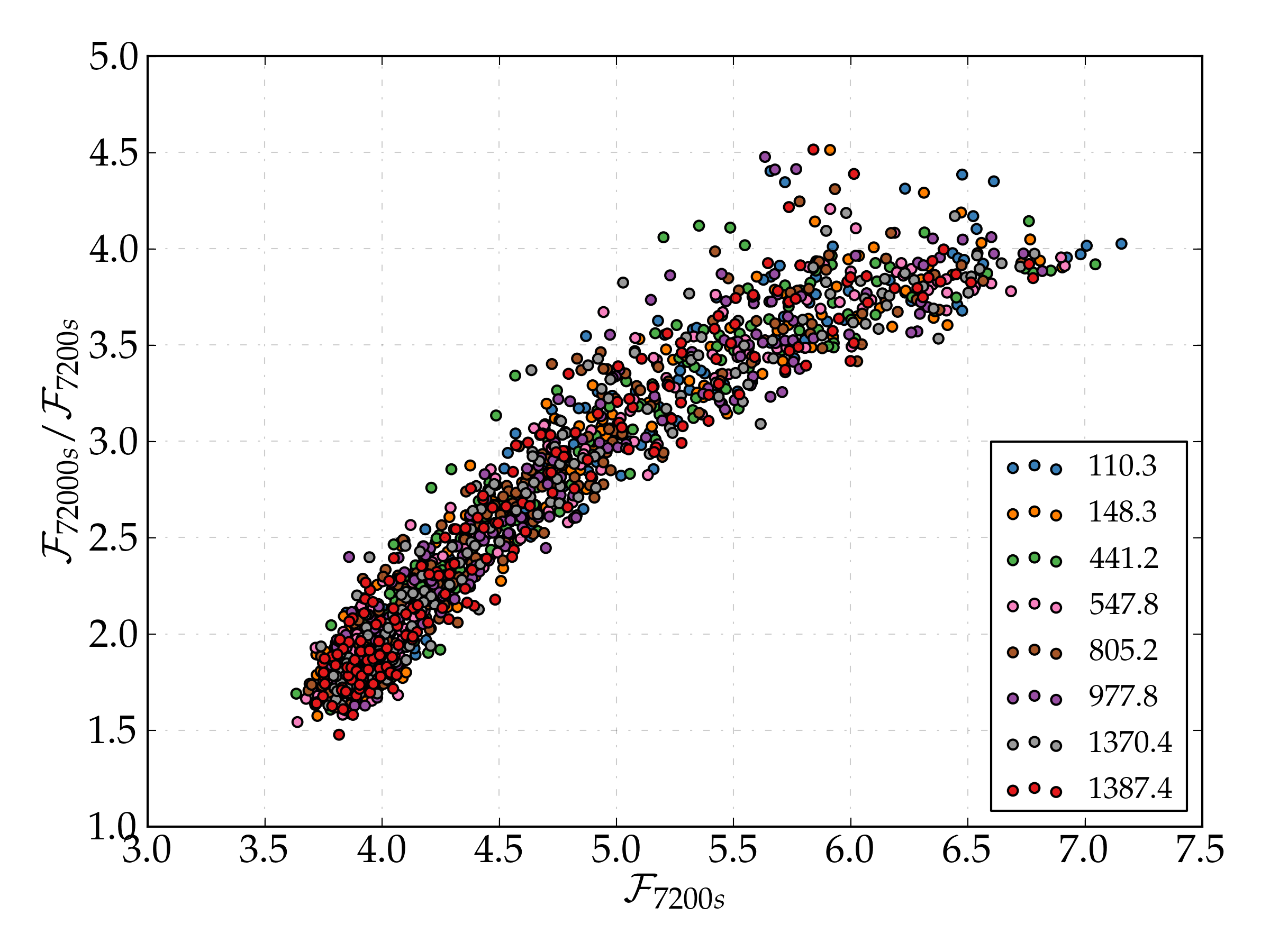}
\includegraphics[width=1.0\columnwidth]{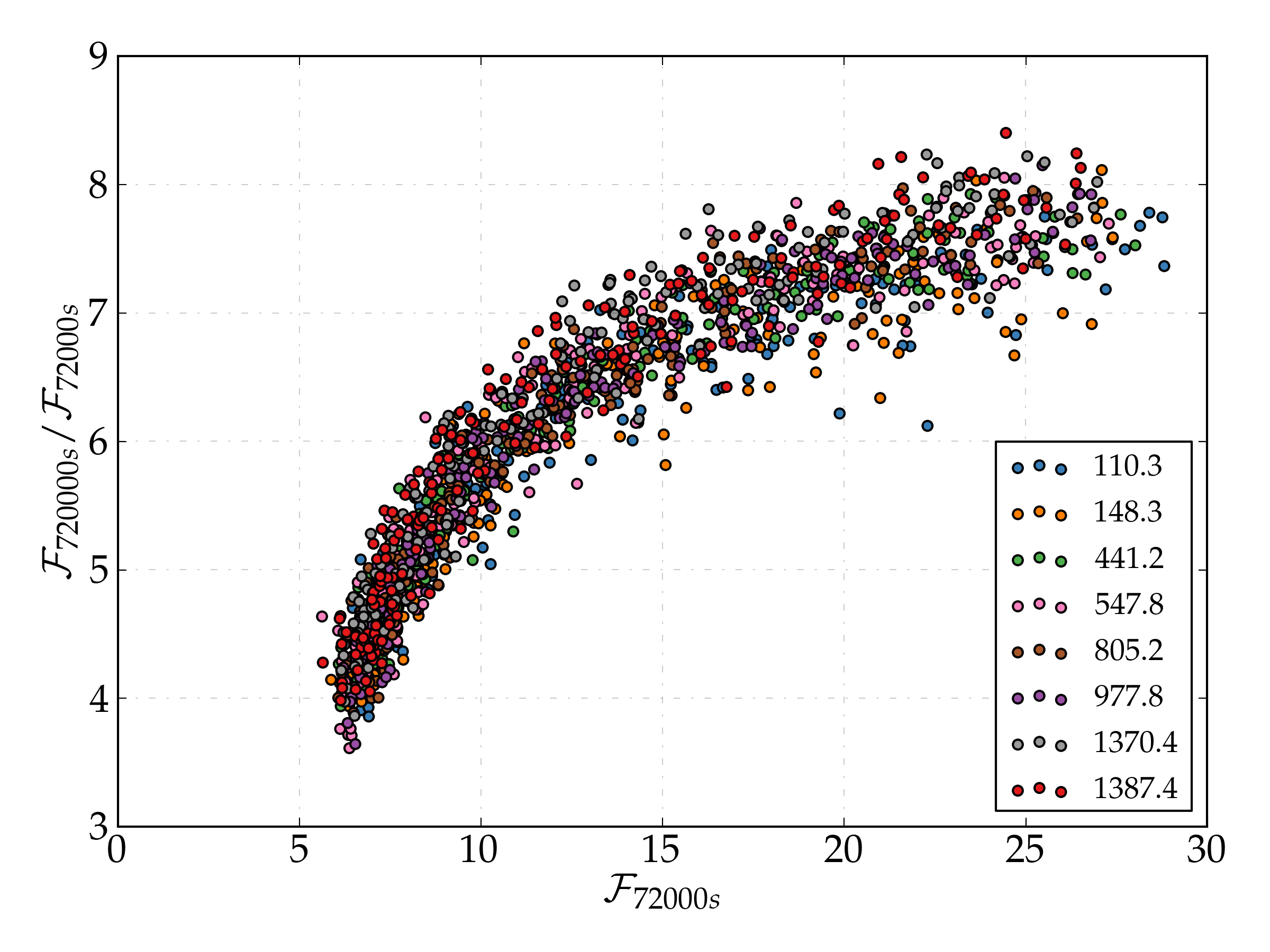}
\caption{Results from software injections for the first (left panel) and second (right panel) comparisons between the first and last two stages of Table \ref{tab:SkyHoughFolInj} of the \textit{SkyHough} follow-up. The vertical axis shows the quotient between the top candidates at the two stages, and the horizontal axis shows the values of the top candidates at the stage with lowest $T_c$. The lowest points, equal to 1.47 and 3.66 respectively, set the thresholds for the follow-up veto. Each color represents a different frequency band.}
\label{fig:SkyHoughFollowUpInj3}
\end{center}
\end{figure*}

\begin{figure}[tbp]
\includegraphics[width=1.0\columnwidth]{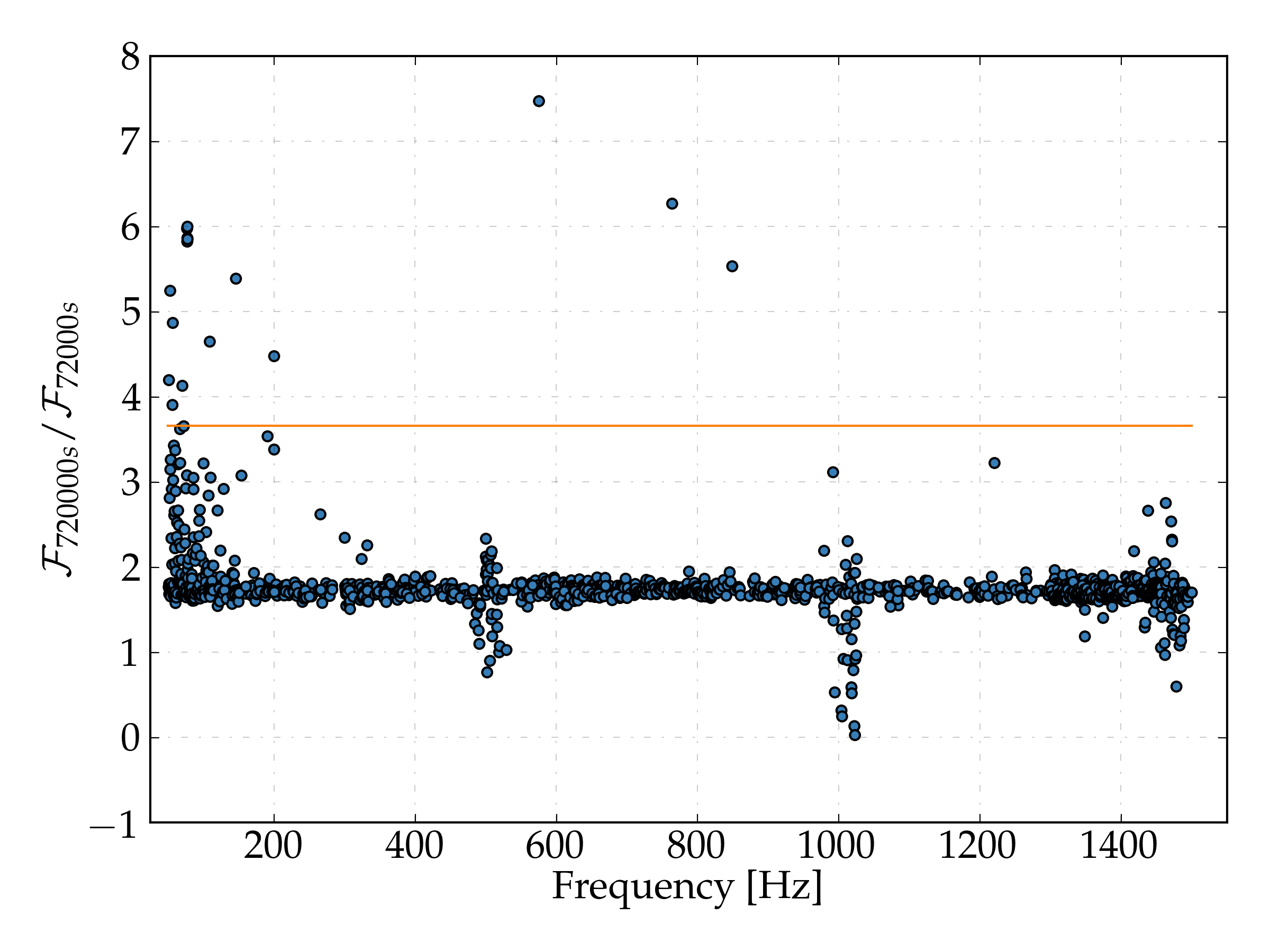}
\caption{Results of the second follow-up comparison (using stages II and III) for the \textit{SkyHough} outliers. Only 17 outliers are above the threshold. The horizontal line marks the threshold at 3.66, which was obtained in Section \ref{SkyHoughFollowup}.}
\label{fig:SkyHoughOutliers2}
\end{figure}

The next step is to apply the $\mathcal{F}$-statistic follow-up method described in Section \ref{SkyHoughFollowup} to these 1539 outliers. The thresholds obtained are 1.47 and 3.66 for the first and second comparison respectively, as shown in Fig \ref{fig:SkyHoughFollowUpInj3}. Only 17 outliers are above the threshold at 3.66, as shown in Fig. \ref{fig:SkyHoughOutliers2}. All of the outliers which are above the final threshold correspond to one of the hardware injections listed in Table \ref{tab:hardwinjections} or to one known source of instrumental noise, listed in \cite{LinesCombs}. The 17 surviving outliers and their parameters are listed in Table \ref{tab:SkyHoughOutliers}, with comments about their likely origin.

We recover 6 of the 9 hardware injections that are in the \textit{SkyHough} searched parameter space. We lose the other three mainly for two reasons: there were brief periods when the hardware injections were not active, which causes the increase of the $\mathcal{F}$-statistic to not be as high as it should be (this happens to two of the three lost hardware injections, which are present in our initial list of 1539 outliers); we only select one cluster per 0.1 Hz band, and if in that band there is a more significant cluster due to a noise disturbance the signal cluster will not be followed (this happens to one of the three lost hardware injections, which forms a cluster but a more significant noise disturbance is present in that 0.1 Hz band).

Although no detections were made, we produce all-sky averaged upper limits on the strain of the signal $h_0$ (these upper limits are valid for all the frequency bands except the ones which have one of the 1539 outliers). We add software simulated signals to the original data by using \textit{lalapps\_Makefakedata\_v5}. We have injected signals at 10 different 0.1 Hz bands for each of the four coherent times (a total of 40 bands), which can be seen in Table \ref{tab:SkyHoughULFreqBands}. These are bands which don't have outliers or instrumental known sources of lines or combs.

\begin{table}[tbp]
\begin{center}
\begin{tabular}{ c c c c }
\hline
$T_c$ & Frequency & Injected $\mathcal{D}$               & $\mathcal{D}^{95\%}$ \\ 
\multicolumn{1}{c}{[s]}   & \multicolumn{1}{c}{[Hz]}  &  [$1/\mathrm{Hz}^{-1/2}$] & [$1/\mathrm{Hz}^{-1/2}$] \\
\hline \hline
\multirow{3}{*}{3600} & 110.3,136.1,148.3, & 27.5, 28.5, & \multirow{3}{*}{32.4} \\
& 165.6,182.6,206.1, & 29.5, 30.5, & \\
& 225.6,241.5,261.3,286.7 & 31.5 & \\ \hline
\multirow{3}{*}{2700} & 311.6,325.4,342.5, & 26.5, 27.5, & \multirow{3}{*}{29.2}  \\
& 363.3,394.0,412.4, &  28.5, 29.5, & \\
& 432.8,441.2,523.4,547.8 & 30.5 & \\ \hline
\multirow{3}{*}{1800} & 594.6,661.1,741.4, & 23.0, 24.0, & \multirow{3}{*}{25.2} \\
& 805.0,866.2,933.1, &  25.0, 26.0, & \\
& 977.6,1064.7,1141.5,1250.7 & 27.0 & \\ \hline
\multirow{3}{*}{900}  & 1313.4,1331.7,1358.4, & 22.0, 23.0 & \multirow{3}{*}{22.3} \\
& 1370.3,1388.8,1402.4, & 24.0, 25.0, & \\
& 1423.1,1430.3,1443.4,1464.6 & 26.0 & \\
\hline
\end{tabular}
\caption{The second column shows the frequency bands used to estimate the \textit{SkyHough} upper limits on gravitational-wave signal amplitude $h_0$. The third column shows the injected sensitivity depth values given by equation \eqref{eq:sensdepth}, and the last column shows the sensitivity depth at $95\%$ confidence for each group.}
\label{tab:SkyHoughULFreqBands}
\end{center}
\end{table}

We have used 5 different sensitivity depths at each coherent time, with 400 signals per sensitivity depth. The sensitivity depth is given by:
\begin{align}
\mathcal{D} = \frac{\sqrt{S_n}}{h_0},
\label{eq:sensdepth}
\end{align}
where $S_n$ is the one-sided power spectral density. We inject signals at random positions in the sky, covering the full spin-down/up range and with random polarization, inclination and initial phase. For each band and depth, we calculate the efficiency (number of detected signals divided by total number of signals). We follow the same procedure as in the all-sky search: we run the main search and then apply coincidences, clustering and the population veto. We assume that a signal is detected if the total distance from the recovered cluster to the actual injection is less than 13 bins.

\begin{figure}[tbp]
\includegraphics[width=1.0\linewidth]{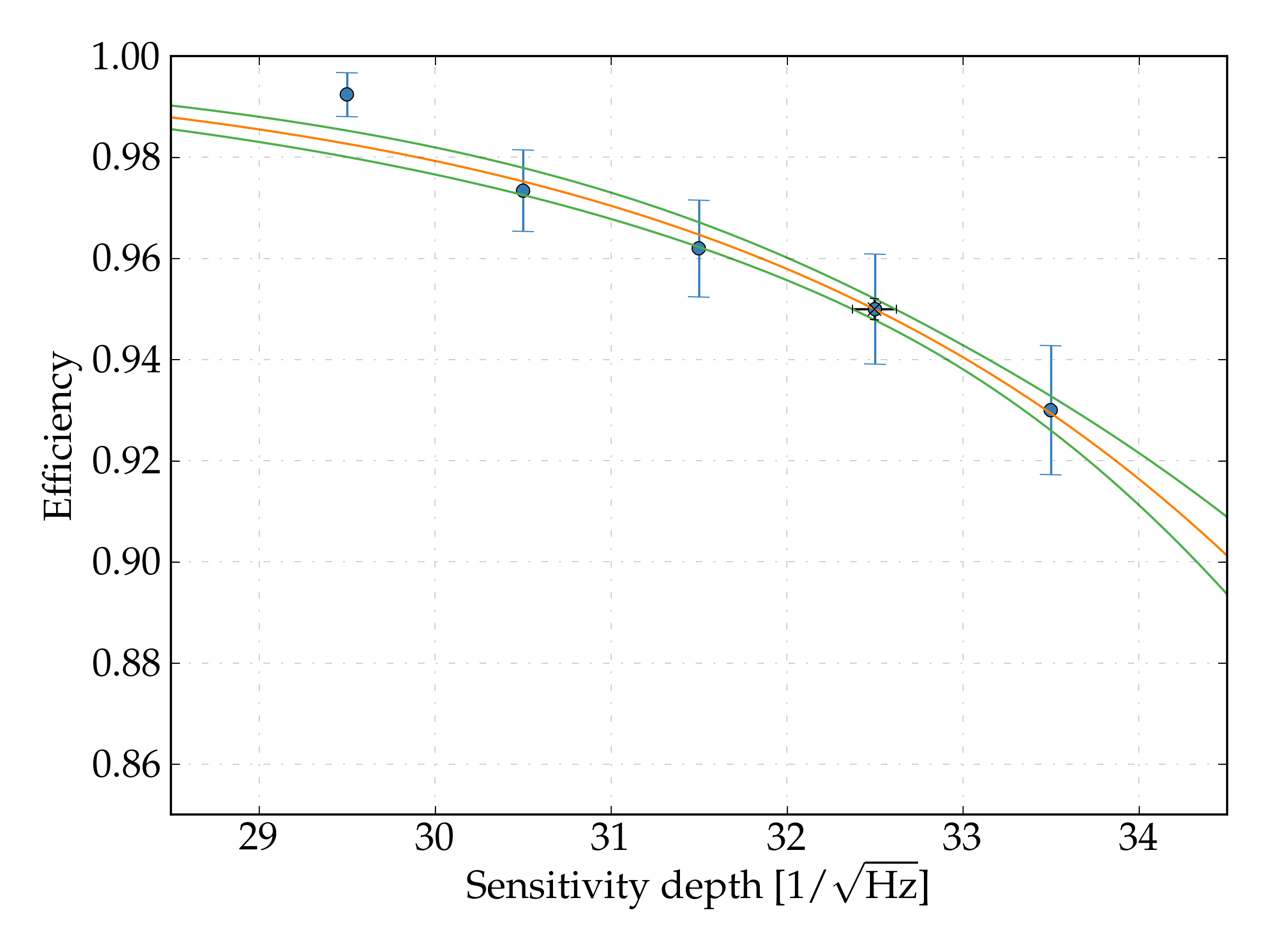}
\caption{Efficiency as a function of sensitivity depth and fitting at 148.3 Hz for the \textit{SkyHough} pipeline. The vertical error bars for the blue points show the 1-sigma binomial error. The 95\% efficiency point (indicated with a black cross) also shows a 1-sigma error bar, calculated with equation \eqref{eq:SkyHoughErrorFit}.}
\label{fig:SkyHoughULFitting}
\end{figure}

At each of the 40 frequency bands, we fit a sigmoid given by:
\begin{align}
s= 1 - \frac{1}{1+e^{b(x-a)}}.
\end{align}
An example of this fitting can be seen in Fig. \ref{fig:SkyHoughULFitting}. From the estimated coefficients $a$ and $b$ along with the covariance matrix $C_{ab}$, we calculate the 1-sigma envelope (the error) on the fit, which is given by:
\begin{align}
\sigma_S = \pm \sqrt{\left(\frac{\delta s}{\delta a}\right)^2 C_{aa} + \left(\frac{\delta s}{\delta b}\right)^2 C_{bb} + 2\frac{\delta s}{\delta a}\frac{\delta s}{\delta b} C_{ab}}.
\label{eq:SkyHoughErrorFit}
\end{align}

\begin{figure}[tbp]
\includegraphics[width=1.0\linewidth]{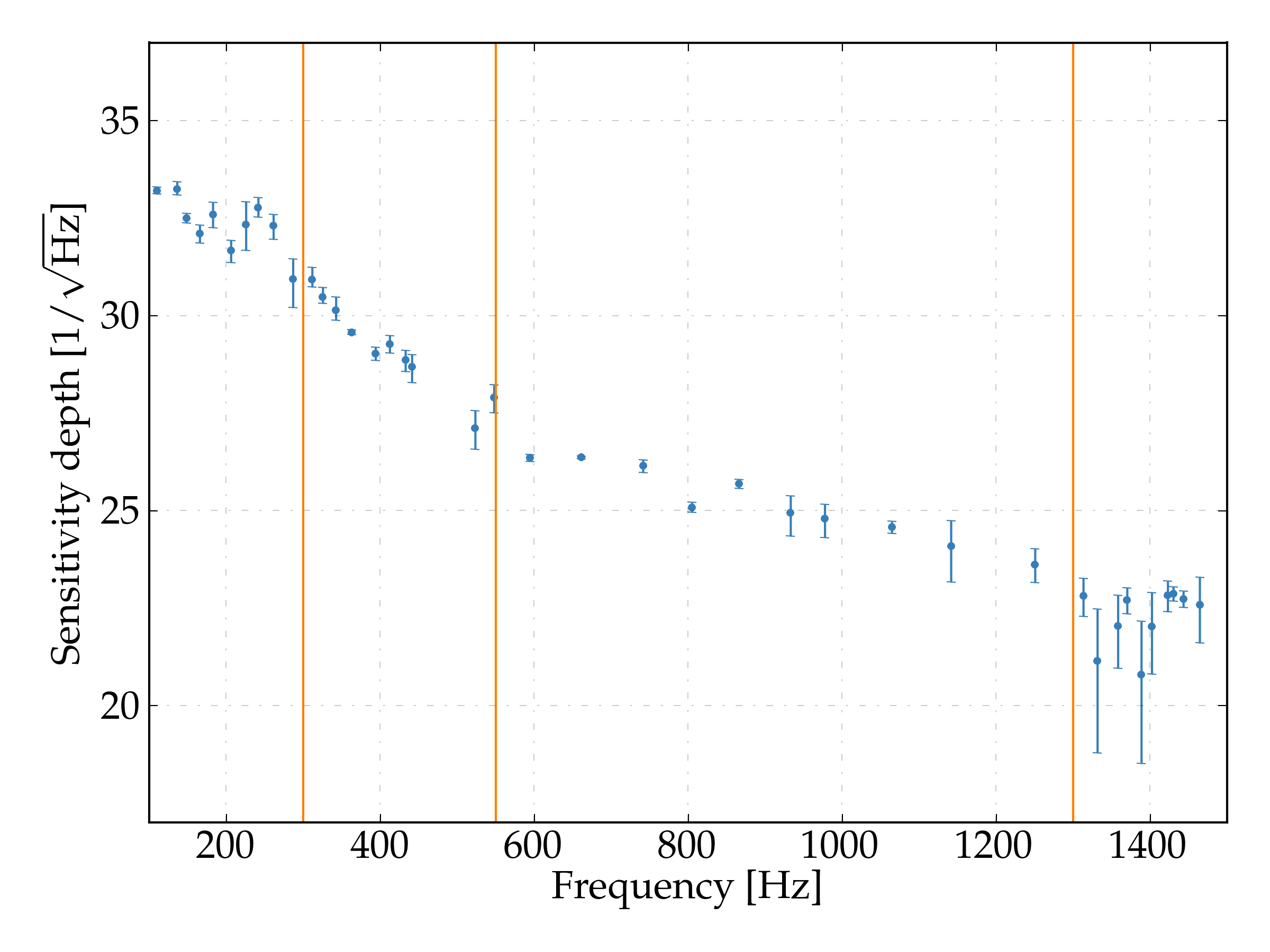}
\caption{The 95\% sensitivity depths at each of the 40 frequency bands, with a 1-sigma error bar, for the \textit{SkyHough} pipeline. The three vertical lines separate the four regions with different coherent time (3600 s, 2700 s, 1800 s and 900 s).}
\label{fig:SkyHoughSensDepth}
\end{figure}

\begin{figure}[tbp]
\includegraphics[width=1.0\columnwidth]{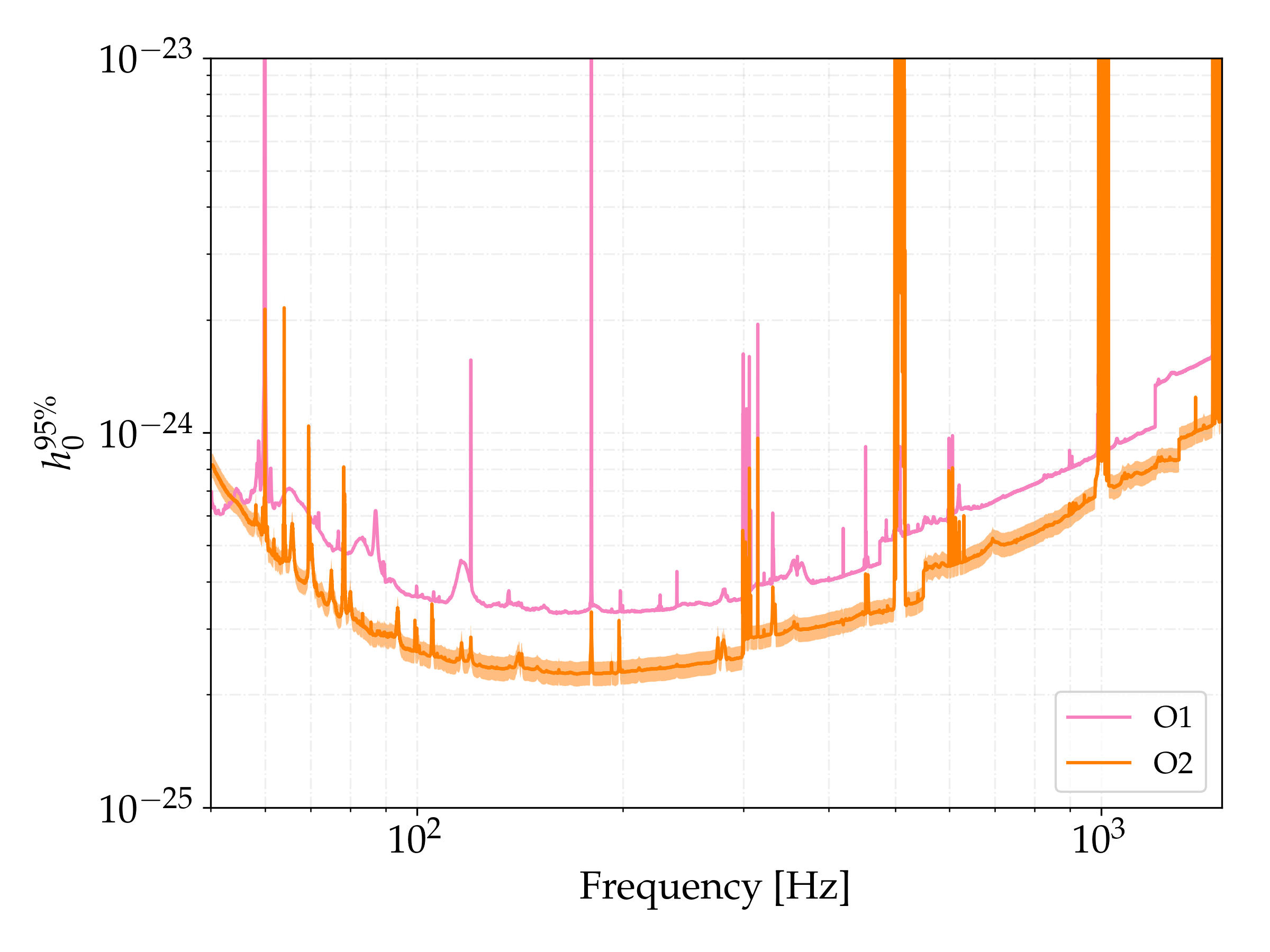}
\caption{$95\%$ upper limits on $h_0$ for the \textit{SkyHough} pipeline. The orange trace shows the results for the O2 search, with a shadow enclosing a 7.5\% error obtained from Fig. \ref{fig:SkyHoughSensDepth}, while the pink trace shows the results obtained in the O1 search. These results are valid for all frequncy bands except the 1539 bands where one outlier is present.}
\label{fig:SkyHoughResultsUL}
\end{figure}

\begin{figure}[tbp]
  \begin{center}
  \includegraphics[width=\columnwidth]{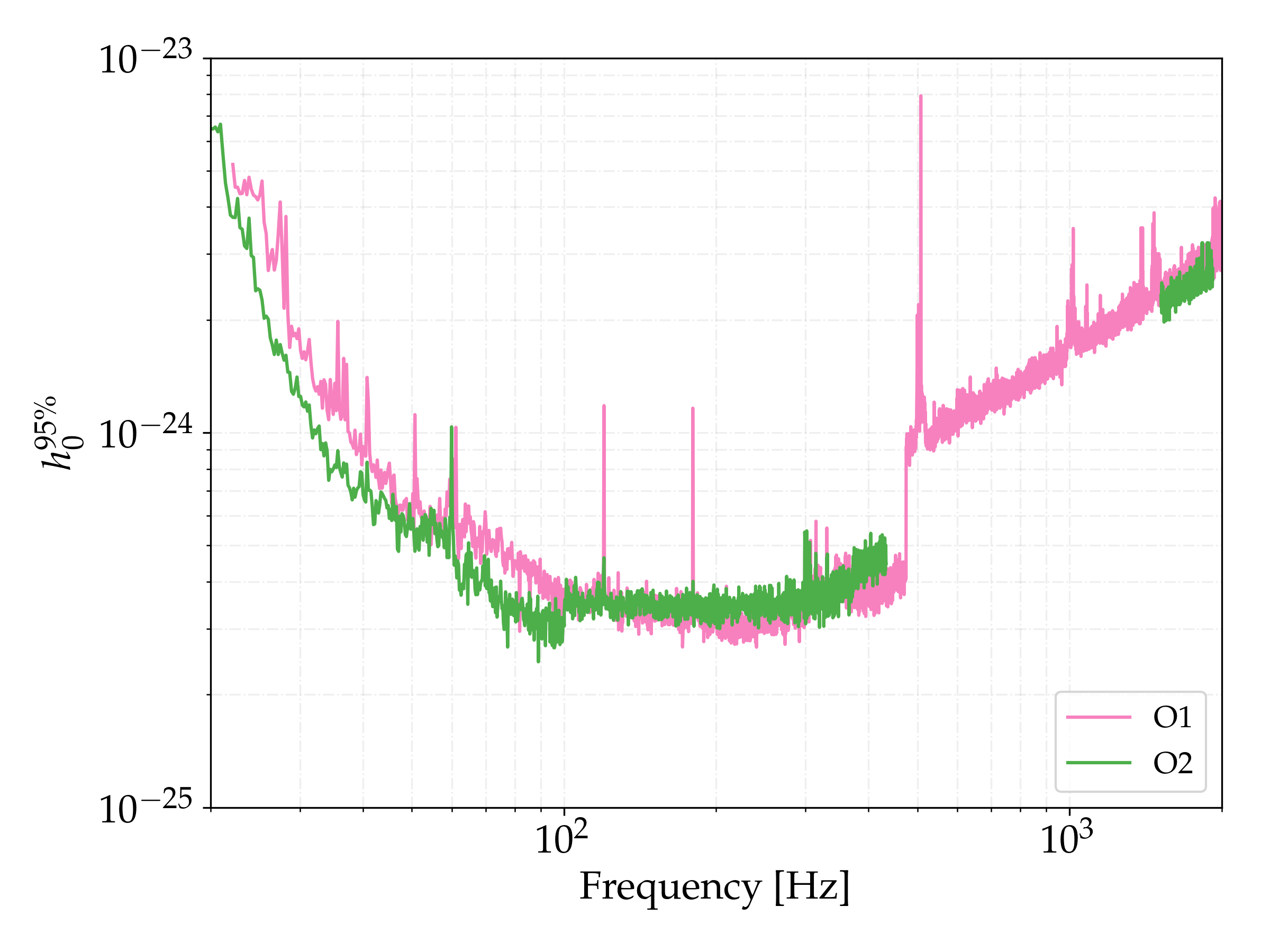}
  \caption{Comparison of O1 and O2 95\% upper limits on $h_o$ for the {\td} pipeline.}
  \label{fig:tdfstat_O2_upper_limits}
  \end{center}
\end{figure}

After finding the $95\%$ efficiency sensitivity depth at each of the 40 frequency bands (which can be seen in Fig. \ref{fig:SkyHoughSensDepth}), we calculate a mean sensitivity depth for each of the four different frequency regions. The results are given in Table \ref{tab:SkyHoughULFreqBands}. From these results and by using equation \eqref{eq:sensdepth}, we calculate the upper limits on $h_0$, which are shown in Fig. \ref{fig:SkyHoughResultsUL}. The trace has a shadow enclosing a 7.5\% error, which we obtain by estimating the maximum difference in each of the four frequency regions shown in Fig. \ref{fig:SkyHoughSensDepth} between the 10 different points and the mean sensitivity depth. Fig. \ref{fig:SkyHoughResultsUL} also shows a comparison with the results obtained in the previous search with O1 data.

\subsection{Time-Domain {\Fstat} results}
\label{sec:TDFstatResults}

In the $[20$-$100]$~Hz, $[100$-$434]$~Hz, and $[1518$-$1922]$~Hz bandwidth
ranges under study, 3380 0.25-Hz wide sub-bands in total were analyzed. As a
results of vetoing candidates around the known interference lines, a certain
fraction of the bandwidth was not analyzed.  As a result $16.0$\% of the band
under study was vetoed.

\begin{figure}[tbp]
  \includegraphics[width=\columnwidth]{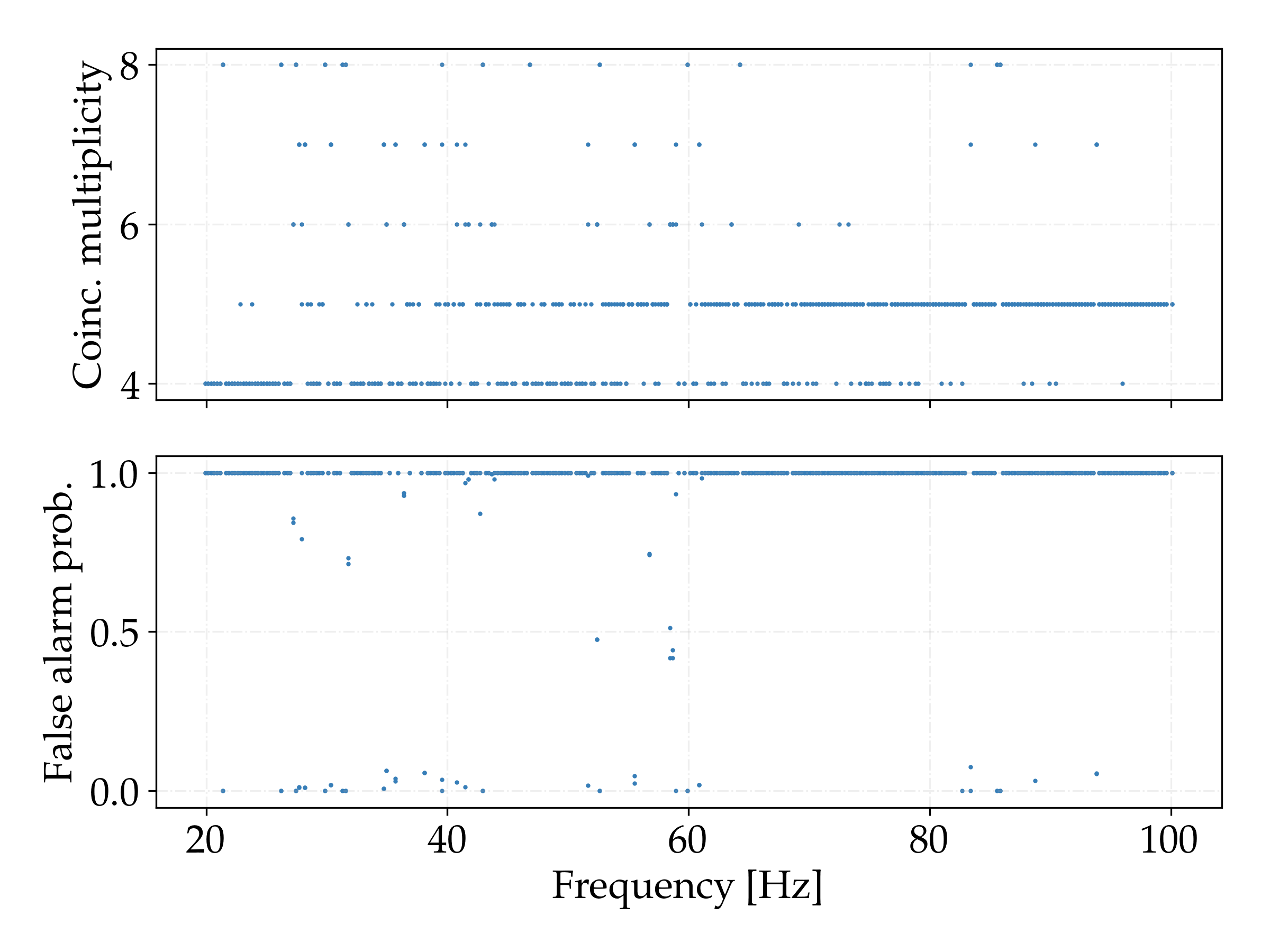}
  \caption{Results of {\td} pipeline coincidences as for 
frequency band of $[20$-$100]$~Hz. Top panel: maximum coincidence multiplicity. Bottom panel:
false alarm probability for the coincidence with the maximum multiplicity.}
  \label{fig:tdfstat_O2_Coinc_FA_24d}
\end{figure}

\begin{figure}[tbp]
  \includegraphics[width=\columnwidth]{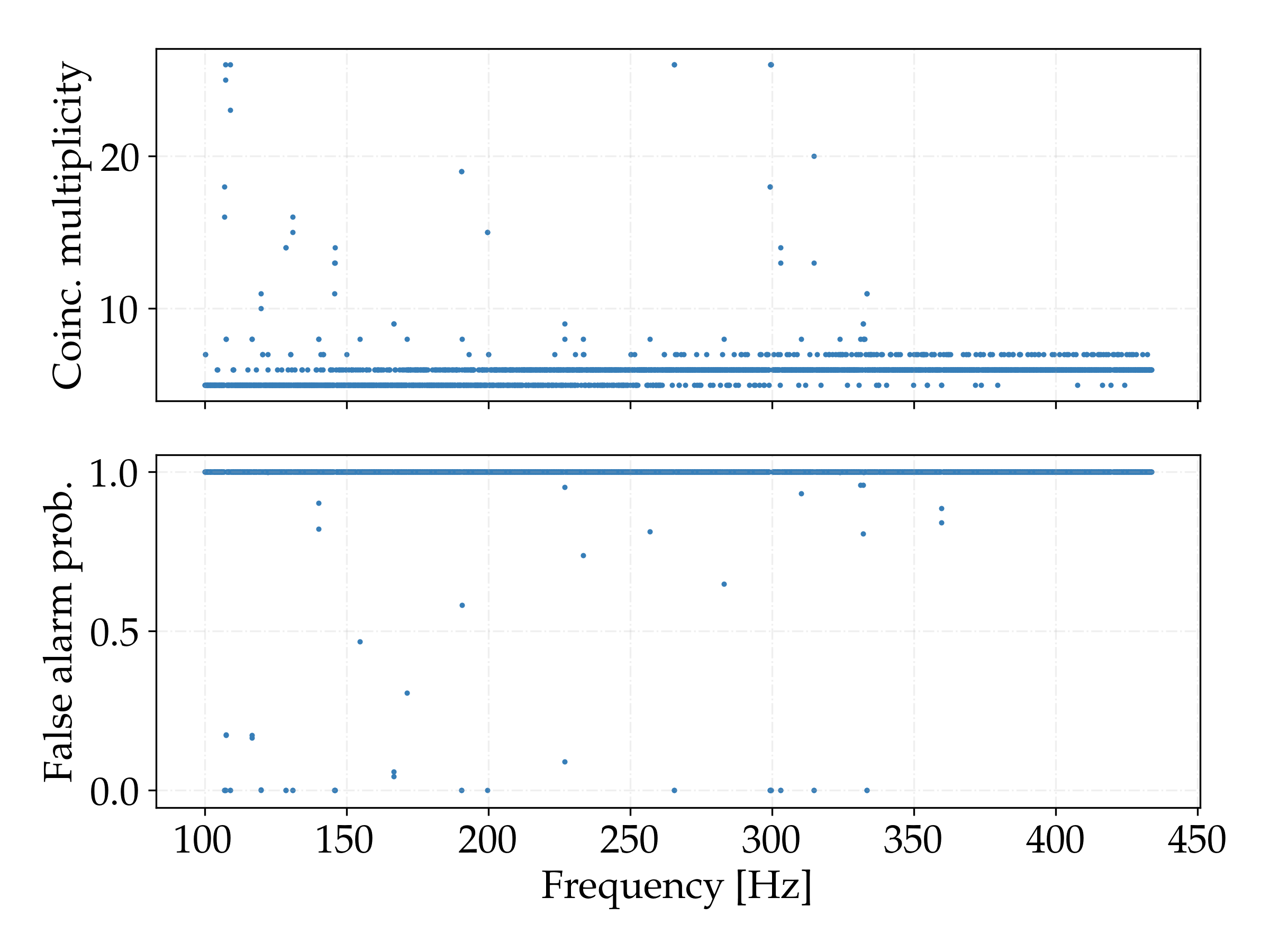}
  \caption{Results of {\td} pipeline coincidences as for 
frequency band of $[100$-$434]$~Hz. Top panel: maximum coincidence multiplicity. Bottom panel:
false alarm probability for the coincidence with the maximum multiplicity.}
  \label{fig:tdfstat_O2_Coinc_FA_6d}
\end{figure}

\begin{figure}[tbp]
  \includegraphics[width=\columnwidth]{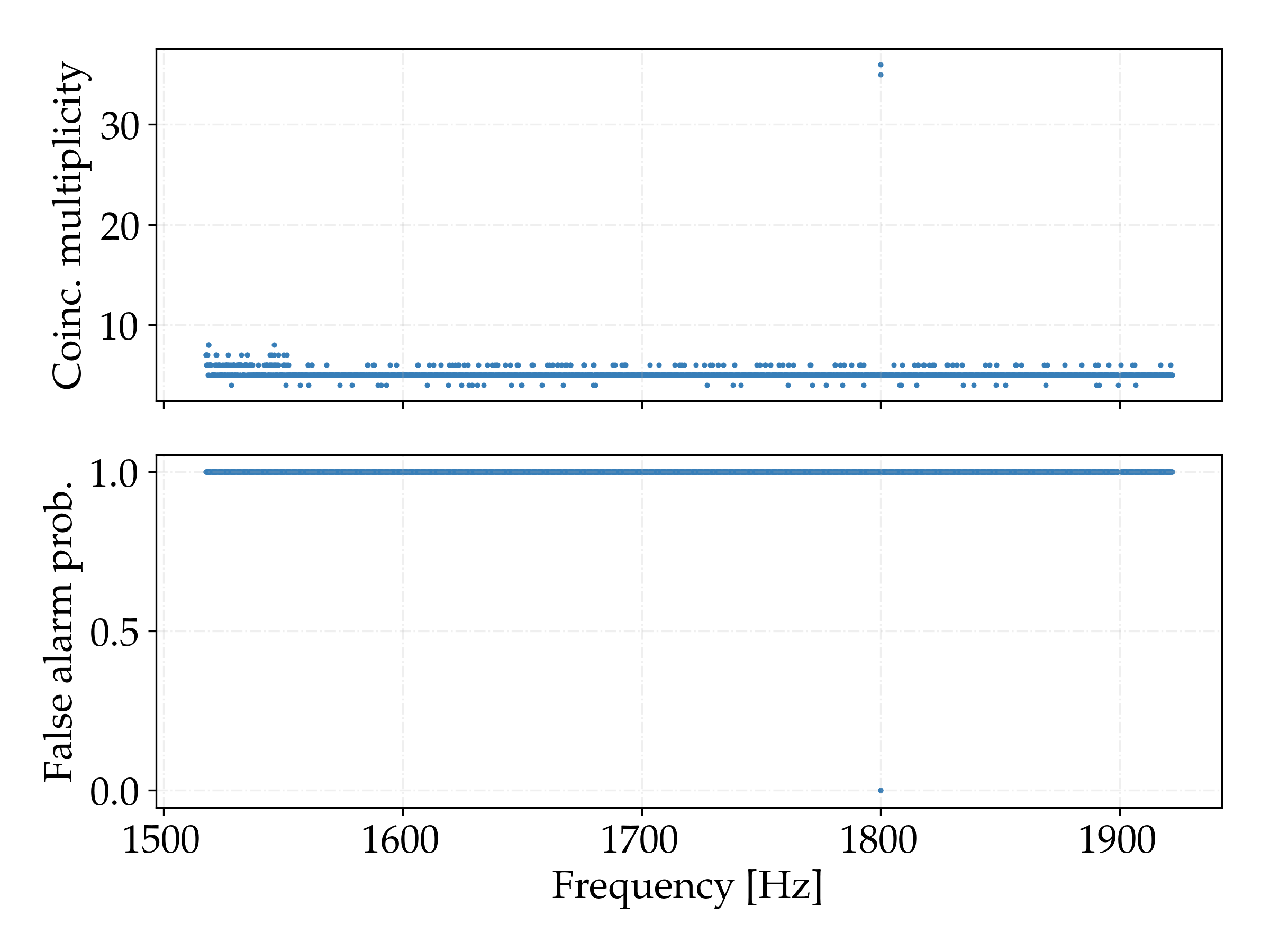}
  \caption{Results of {\td} pipeline coincidences as for 
frequency band of $[1518$-$1922]$~Hz. Top panel: maximum coincidence multiplicity. Bottom panel:
false alarm probability for the coincidence with the maximum multiplicity.}
  \label{fig:tdfstat_O2_Coinc_FA_2d}
\end{figure}

In Figs.~\ref{fig:tdfstat_O2_Coinc_FA_24d}, \ref{fig:tdfstat_O2_Coinc_FA_6d}
and \ref{fig:tdfstat_O2_Coinc_FA_2d} the results of the coincidence search are
presented for bandwidth ranges $[20$-$100]$~Hz,  $[100$-$434]$~Hz, and
$[1518$-$1922]$~Hz respectively. The top panels show the maximum coincidence
multiplicity for each of the sub-bands analyzed. The maximum multiplicity is an
integer that varies from 4 to the number of time frames in each of the
bandwidth analyzed. This is because we record coincidences of multiplicity of
at least 4. 


The bottom panel of Figs.~\ref{fig:tdfstat_O2_Coinc_FA_24d},
\ref{fig:tdfstat_O2_Coinc_FA_6d} and \ref{fig:tdfstat_O2_Coinc_FA_2d} shows the
results for the false alarm probability of coincidence for the coincidence with
the maximum multiplicity. This is the probability that a coincidence among candidates from all the time-domain segments
in a given frequency sub-band occurs by chance. This false alarm probability is calculated using the formula (A.6)
in the Appendix of \cite{VSR1TDFstat}.

We define outliers as those coincidences with false alarm probabilities less
than $0.1$\% This criterion was adopted in our Virgo data
search~\cite{VSR1TDFstat} and also in one of the {\EatH} searches 
\cite{S4CWAllSky}. As a result we obtained 30 outliers.  Among the 30
outliers, 8 are identified with the hardware injections.  Table
\ref{tab:tdfstat_hi} presents the estimated parameters obtained for these
hardware injections, along with the absolute errors of the reconstructed
parameters (the differences with respect to the injected parameters).

The remaining 23 outliers are listed in Table \ref{tab:TDFstat_outliers}.  They
include 16 that are seen only in H1 data and 3 in only the L1 data. In the case
of the remaining outliers their amplitude is stronger in H1 than in L1 whereas the
noise level in H1 is lower than in L1. Consequently no credible gravitational
wave candidates were found.

\section{Conclusions}
In this paper we have presented the first results of an all-sky search for CW signals using Advanced LIGO O2 data with three different pipelines, covering a frequency range from 20 to 1922 Hz and a first frequency derivative from $-1\times10^{-8}$ to $2\times10^{-9}$ Hz/s. For this broad range in parameter space, this is the most sensitive search up to 1500 Hz. Each search found many outliers which were followed-up but none of them resulted in a credible astrophysical CW signal. On the contrary, they were ascribable to noise disturbances, to hardware injections, or consistent with noise fluctuations. 

Although no detections have been made, we have placed interesting 95\% CL upper limits on the gravitational wave strain amplitude $h_0$, the most sensitive being $\simeq 1.7\times10^{-25}$ in the 123-124 Hz region, as shown in Fig. \ref{fig:globalUL}. The improved results over the O1 search are due to the better sensitivity of the detectors, the use of a longer dataset and improvements of the pipelines. For the semi-coherent methods used in this analysis, the strain sensitivity is proportional to \cite{Prix,Shaltev,EstimatingSensitivity}: 
\begin{align}
        h_0 \propto \sqrt{\frac{S_n}{T_{coh}}} N^{\sim 1/4} .
        \label{eq:sensi}
\end{align}
For the same observing time, an improvement of 1.15 in the noise floor of both detectors translates to an expected improvement of 1.15 of the minimum detectable strain (looking at Fig. 1, we see that H1 has improved by ${\sim}1.15$, while L1 has improved less, so at most a factor of 1.15 is what we expect from the improved noise floors). To ascertain the influence of the run duration (assuming that the same coherent time has been used), we compare the different number of segments used. Comparing O1 \cite{O1CWAllSkyFull} where ${\sim}6600$ segments were used by the SkyHough pipeline (similar numbers apply also for the other pipelines), with O2 where ${\sim}10600$ segments have been used, the expected improvement is of ${\sim}1.12$, which multiplied by 1.15 results in a ${\sim}1.28$ factor. The difference between this and the 1.4 observed improvement is due to the enhancement of the pipelines.

By converting the upper limits to an astrophysical reach, as shown in Fig. \ref{fig:AstroReach}, we see that the searches presented in this paper provide already astrophysical interesting results. For instance, in the ``bucket'' region (around $\simeq 150$ Hz), we would be able to detect a CW signal from a neutron star within a distance of 100 pc if its ellipticity were at least $10^{-6}$. Similarly, in the middle frequency range, around $\simeq 500$ Hz, we would be able to detect the CW signal up to a distance of 1 kpc, with $\epsilon > 10^{-6}$. Finally at higher frequencies, around $\simeq 1500$ Hz, the same signal would be  detectable up to a distance of 10 kpc if  $\epsilon > 10^{-6}$ and 1 kpc if  $\epsilon > 10^{-7}$.
Such levels of ellipticity are comparable or below the maximum value we may expect for neutron stars described by a standard equation of state \cite{crust_limit2}. Although approximately 2500 neutron stars have been observed through their electro-magnetic emission, a much larger number of undiscovered neutron stars is expected to exist in our Galaxy, a small fraction of which could be nearby.

Further all-sky analyses are planned on O2 data, by extending the parameter space and looking at sub-threshold candidates. The O3 observing run has started in April 2019 and will last for approximately 1 year. The full network of LIGO and Virgo detectors is being upgraded and improved, and we expect that the noise floor in O3 run will be significantly better than for O2. This, and the foreseen longer run duration, will make future searches more sensitive (if the noise floor improves by 1.5 and the run is 1.5 times longer than O2, we expect an improvement of ${\sim}1.66$ on the strain upper limits), increasing the chances of a CW detection or allowing us to place tighter constraints on the non-asymmetries of neutron stars in our galaxy and to put constraints on the unseen neutron star population.

\begin{acknowledgments}
The authors gratefully acknowledge the support of the United States
National Science Foundation (NSF) for the construction and operation of the
LIGO Laboratory and Advanced LIGO as well as the Science and Technology Facilities Council (STFC) of the
United Kingdom, the Max-Planck-Society (MPS), and the State of
Niedersachsen/Germany for support of the construction of Advanced LIGO 
and construction and operation of the GEO600 detector. 
Additional support for Advanced LIGO was provided by the Australian Research Council.
The authors gratefully acknowledge the Italian Istituto Nazionale di Fisica Nucleare (INFN),  
the French Centre National de la Recherche Scientifique (CNRS) and
the Foundation for Fundamental Research on Matter supported by the Netherlands Organisation for Scientific Research, 
for the construction and operation of the Virgo detector
and the creation and support of the EGO consortium. 
The authors also gratefully acknowledge research support from these agencies as well as by 
the Council of Scientific and Industrial Research of India, 
the Department of Science and Technology, India,
the Science \& Engineering Research Board (SERB), India,
the Ministry of Human Resource Development, India,
the Spanish Agencia Estatal de Investigaci\'on,
the Vicepresid\`encia i Conselleria d'Innovaci\'o, Recerca i Turisme and the Conselleria d'Educaci\'o i Universitat del Govern de les Illes Balears,
the Conselleria d'Educaci\'o, Investigaci\'o, Cultura i Esport de la Generalitat Valenciana,
the National Science Centre of Poland,
the Swiss National Science Foundation (SNSF),
the Russian Foundation for Basic Research, 
the Russian Science Foundation,
the European Commission,
the European Regional Development Funds (ERDF),
the Royal Society, 
the Scottish Funding Council, 
the Scottish Universities Physics Alliance, 
the Hungarian Scientific Research Fund (OTKA),
the Lyon Institute of Origins (LIO),
the Paris \^{I}le-de-France Region, 
the National Research, Development and Innovation Office Hungary (NKFIH), 
the National Research Foundation of Korea,
Industry Canada and the Province of Ontario through the Ministry of Economic Development and Innovation, 
the Natural Science and Engineering Research Council Canada,
the Canadian Institute for Advanced Research,
the Brazilian Ministry of Science, Technology, Innovations, and Communications,
the International Center for Theoretical Physics South American Institute for Fundamental Research (ICTP-SAIFR), 
the Research Grants Council of Hong Kong,
the National Natural Science Foundation of China (NSFC),
the Leverhulme Trust, 
the Research Corporation, 
the Ministry of Science and Technology (MOST), Taiwan
and
the Kavli Foundation.
The authors gratefully acknowledge the support of the NSF, STFC, MPS, INFN, CNRS, INFN-CNAF, PL-Grid, and the
State of Niedersachsen/Germany for provision of computational resources. 
Work at SURFsara and Nikhef has been performed using resources of the
Dutch e-Infrastructure, which is financially supported by the Nederlandse
Organisatie voor Wetenschappelijk Onderzoek (Netherlands Organisation for
Scientific Research, NWO) and the Dutch higher education and research partnership
for network services and information and communication technology (SURF).
This article has LIGO document number LIGO-P1900012-v7.

\end{acknowledgments}

\section{Appendix}

\begin{table}[tbp]
\begin{center}
\begin{tabular}{ r r r r r}
\hline
Label & \multicolumn{1}{c}{Frequency} & \multicolumn{1}{c}{Spin-down} & \multicolumn{1}{c}{$\alpha$} & \multicolumn{1}{c}{$\delta$} \\
 & \multicolumn{1}{c}{[Hz]} & \multicolumn{1}{c}{[nHz/s]} & \multicolumn{1}{c}{[deg]} & \multicolumn{1}{c}{[deg]} \\
\hline \hline
ip0   &  265.575533  & $-4.15 \times 10^{-3}$    &   71.55193     &  -56.21749 \\
ip1   &  848.969641  & $-3.00 \times 10^{-1}$    &   37.39385     &  -29.45246 \\
ip2   &  575.163521  & $-1.37 \times 10^{-4}$    &  215.25617     &    3.44399 \\
ip3   &  108.857159  & $-1.46 \times 10^{-8}$    &  178.37257     &  -33.4366  \\
ip4   & 1393.540559  & $-2.54 \times 10^{-1}$    &  279.98768     &  -12.4666  \\
ip5   & 52.808324    & $-4.03 \times 10^{-9}$    &  302.62664     &  -83.83914 \\
ip6   &  146.169370  & $-6.73 \times 10^{0}$     &  358.75095     &  -65.42262 \\
ip7   & 1220.555270  & $-1.12 \times 10^{0}$     &  223.42562     &  -20.45063 \\
ip8   &  191.031272  & $-8.65 \times 10^{0}$     &  351.38958     &  -33.41852 \\
ip9   &  763.847316  & $-1.45 \times 10^{-8}$    &  198.88558     &   75.68959 \\
ip10  &   26.341917  & $-8.50 \times 10^{-2}$    &  221.55565     &   42.87730 \\
ip11  &   31.424758  & $-5.07 \times 10^{-4}$    &  285.09733     &  -58.27209 \\
ip12  &   38.477939  & $-6.25 \times 10^{0}$     &  331.85267     &  -16.97288 \\
ip13  &   12.428001  & $-1.00 \times 10^{-2}$    &  14.32394      &  -14.32394 \\
ip14  & 1991.092401  & $-1.00\times 10^{-3}$     & 300.80284      &  -14.32394 \\
\hline
\end{tabular}
\caption[Parameters of hardware injections]{Parameters of the hardware-injected simulated continuous-wave signals during the O2 data run (parameters given at epoch GPS $1130529362$).} 
\label{tab:hardwinjections}
\end{center}
\end{table}

\begin{table*}[tbp]
\begin{center}
\begin{tabular}{ r r r r r r r r r}
\hline
Outlier & \multicolumn{1}{c}{Frequency} & \multicolumn{1}{c}{Spin-down} & \multicolumn{1}{c}{$\alpha$} & \multicolumn{1}{c}{$\delta$} & Population & $\bar{s}_P$ & $\mathcal{F}$ & \multicolumn{1}{c}{Description} \\
index   & \multicolumn{1}{c}{[Hz]}      & \multicolumn{1}{c}{[nHz/s]}  & \multicolumn{1}{c}{[deg]}    & \multicolumn{1}{c}{[deg]}    & & & & \\
\hline \hline
7  &  51.0002  &  $-1.8346 \times 10^{-11}$  &  87.0087  &  -66.0873  & 6542 & 50.16 &  86.0406  & 1 Hz comb at H1 and L1 \\
18  &  52.8083  &  $2.2838 \times 10^{-12}$  &  299.6708  &  -83.3562  & 1433 & 124.12 &  654.9832  & Hardware injection 5 \\
36  &  56.0001  &  $-6.3195 \times 10^{-12}$  &  88.8156  &  -66.4443  & 2765 & 210.17 &  157.7912  & 1 Hz comb at H1 and L1 \\
39  &  56.4957  &  $5.7155 \times 10^{-10}$  &  156.8932  &  -49.1672  & 111 & 9.04 &  72.64532  & 1 Hz comb at H1 and L1 \\
113  &  70.0001  &  $-1.2045 \times 10^{-11}$  &  88.1699  &  -66.1180  & 2980 & 121.85 &  165.2225  & 1 Hz comb at H1 and L1 \\
125  &  72.0001  &  $-5.4952 \times 10^{-12}$  &  89.4121  &  -66.5021  & 3471 & 57.69 &  64.56955  & 1 Hz comb at H1 and L1 \\
149  &  76.6788  &  $-1.7870 \times 10^{-10}$  &  74.5911  &  -59.0142  & 397 & 32.66 &  444.1439  & 0.08843966 Hz comb at H1 \\
150  &  76.9442  &  $-1.8308 \times 10^{-10}$  &  74.6723 &  -58.9216  & 558 & 36.12 &  566.2429  & 0.08843966 Hz comb at H1 \\
151  &  77.1207  &  $-1.4130 \times 10^{-10}$  &  77.8483  &  -59.1510  & 129 & 26.81 &  507.6696  & 0.08843966 Hz comb at H1 \\
152  &  77.2090  &  $-1.3071 \times 10^{-10}$  &  79.0815  &  -59.5503  & 93 & 21.51 &  405.0106  & 0.08843966 Hz comb at H1 \\
153  &  77.3855  &  $-7.9419 \times 10^{-11}$  &  279.3909  &  70.9090  & 150 & 14.78 &  154.7181  & Unknown line at H1 \\
316  &  108.8567  &  $4.3434 \times 10^{-11}$  &  182.2717  &  -29.6472  & 3050 & 72.87 &  275.9235  & Hardware injection 3 \\
485  &  145.9203  &  $-6.7301 \times 10^{-9}$  &  358.7866  &  -65.2887 & 675 & 86.45 &  311.7228  & Hardware injection 6 \\
664  &  199.9977  &  $-1.5070 \times 10^{-11}$  &  89.5203  &  -66.2582 & 1249 & 83.46 &  130.1102  & 99.9987 Hz comb at H1 \\
1629  &  575.1638  &  $-2.9038 \times 10^{-11}$  &  215.4209  &  4.0846  & 636 & 564.51 &  4962.3050  & Hardware injection 2 \\
2303  &  763.8471  &  $1.2199 \times 10^{-11}$  &  198.9249  &  75.6197  & 1798 & 637.28 &  2703.2552  & Hardware injection 9 \\
2584  &  848.9591  &  $-3.4856 \times 10^{-10}$  &  37.3061  &  -28.8880  & 443 & 898.46 &  4473.9817  & Hardware injection 1 \\ 
\hline
\end{tabular}
\caption{17 outliers from the \textit{SkyHough} pipeline which survived the follow-up procedure. All of them can be ascribed to a hardware injection or to a known source of instrumental noise. The ($f_0,\dot{f},\alpha,\delta$) values correspond to the center of the cluster returned by the post-processing stage. The $\bar{s}_P$ column shows the mean power significance of the cluster, while $\mathcal{F}$ column shows the $\mathcal{F}$-statistic mean over segments of the top candidate obtained at the last stage of the follow-up. The reference time for these parameters is 1167545839 GPS.}
\label{tab:SkyHoughOutliers}
\end{center}
\end{table*}

\subsection{Frequency Hough hardware injections recovery} 
\label{app:FHHI}
Table \ref{tab:fh_hi} shows the parameters of 
the recovered signals, together with the error with respect to the injected signals. 
\begin{table*}[tbp]
\begin{center}
\begin{tabular}{r r r r r r r}\hline
Label & CR & Frequency [Hz] & Spin-down [nHz/s] & $\alpha$ [deg] & $\delta$ [deg] \\
\hline \hline
ip13 & 27.8 & 12.427537 ($-$0.000008) & $-$0.0070 (0.0030) & 11.14 ($-$0.23) & 17.78 (2.29) \\
ip10 & 147.8 & 26.338028 (-0.000012) & $-$0.08423 (0.00077) & 221.14 (0.043) & 42.79 ($-$0.15)  \\
ip11 & 130.8 & 31.424762 (0.000027) & $-$0.0009 (0.0014) & 284.52 (0.28) & $-$58.27 (0.04) \\
ip12  & 78.7 & 38.192818 ($-$0.000003) &$-$6.2542 ($-$ 0.0042) & 336.06 (2.24) & $-$25.74 ($-$4.35) \\
ip5 & 161.9 & 52.808286($-$0.000038) & 0.00000 ($4 \times 10^{-9}$) & 301.87 ($-$0.37) & $-$83.641 (0.099) \\
ip3 & 67.42 & 108.857166 ($-$0.000007) & $-$0.00088 ($-$ 0.00088) & 178.29 ($-$0.0.2) & $-$33.59 ($-$0.09) \\
ip6 & 76.4 & 145.862390 (0.000036) & $-$6.7245 (0.0055) & 358.85 (0.27) & $-$65.63 ($-$0.01) \\
ip8 & 72.7 & 190.636613 ($-$0.000055) & $-$8.661 ($-$0.011) & 351.20 ($-$0.08) & $-$33.04 (0.19) \\
ip0 & 293.6 & 265.575312 ($-$ 0.000032) & 0.0000 (0.0041) & 71.64 (0.05) & $-$56.29 ($-$0.03) \\
ip2 & 297.3 & 575.163534 (0.000019) & 0.00000 (0.00014) & 215.18 (0.03) & 3.44 ($-$0.02) \\
ip9 & 394.2 & 763.847307 ($-$ 0.000010) & 0.0053 (0.0053) & 198.86 ($-$0.01) & 75.65 ($-$0.02) \\
ip1 & 408.4 & 848.955908 ($-$ 0.000048) & $-$0.2974 (0.0026) & 37.43 (0.02) & $-$29.49 ($-$0.02) \\
 \hline
\end{tabular}
\caption[]{Hardware injection recovery with the \fh\ pipeline. The reported values have been obtained at the end of the full analysis, including the 
follow-up. The values in parentheses are the absolute errors, that is the difference with respect to the injection parameters. The reference time is MJD 57856.826840. The sky position is given in equatorial coordinates.}
\label{tab:fh_hi}
\end{center}
\end{table*}
As shown in the Table, we have been able to detect all the 13 injections done in the analyzed frequency band and the estimated parameters do show  a very good agreement with the injected ones.

\subsection{Frequency Hough selected sub-threshold candidates} 
\label{app:FHST}
 Table \ref{tab:st} shows the parameters of the candidates which have been excluded because, after the follow-up and the verification stages, are in coincidence (within the standard distance, equal to 3) but below the CR threshold value $\mathrm{CR_{\text{thr}}}=7.42$.
\begin{table*}[tbp]  
 \begin{center} 
 \begin{tabular}{c c c c c c} \hline 
 Order number & CR & Frequency [Hz] & Spin-down [nHz/s] & $\alpha$ [deg] & $\delta$ [deg] \\
 \hline \hline 
1 & 4.75 & 184.0774 & $-$5.0337 & 281.0392 & $-$14.3215 \\ 
2 & 4.63 & 294.6539 &  1.9496 & 314.7224 & $-$1.3227 \\ 
3 & 5.02 & 316.2952 & $-$5.0802 &  237.2929 & 22.1560 \\ 
4 & 4.46 & 339.3438 & $-$2.9499 &  181.4295 & 33.5115 \\ 
5 & 4.09 & 353.2267 & $-$5.8594 & 346.0884 & $-$27.8214 \\ 
6 & 4.17 & 377.1483 & $-$9.9569 & 261.5479 & $-$33.9948 \\ 
7 & 3.88 & 403.3479 &  1.8908 & 284.5725 & $-$2.8764 \\ 
8 & 4.29 & 404.0677 & $-$3.3008 & 132.9264 & $-$8.0342 \\ 
9 & 4.11 & 423.3808 & $-$0.0456 &  140.3369 & 83.1005 \\ 
10 & 4.04 & 433.9027 &  0.2334 & 207.0290 & $-$63.3601 \\ 
11 & 4.26 & 457.7948 & $-$2.1637 &  250.2473 & 12.2122 \\ 
12 & 4.18 & 297.8762 & $-$5.9050 & 116.5559 & $-$37.5740 \\ 
13 & 3.75 & 407.0068 &  0.8607 &  308.1622 & 41.2725 \\ 
14 & 4.56 & 456.3493 & $-$2.7138 &  270.8929 & 45.9713 \\ 
15 & 3.59 & 475.5485 & $-$7.0957 & 79.9163 & $-$30.3946 \\ 
16 & 4.60 & 475.7466 & $-$1.0713 & 43.7284 & $-$57.1093 \\ 
17 & 3.18 & 611.3748 &  1.7294 & 111.7053 & $-$31.0260 \\ 
18 & 3.97 & 636.4725 & $-$0.7449 &  161.4437 & 26.4633 \\ 
19 & 3.33 & 636.4718 & $-$0.8265 &  161.1148 & 27.2096 \\ 
20 & 3.50 & 709.5682 & $-$0.7265 & 75.0639 & $-$57.6713 \\ 
21 & 4.21 & 728.2990 &  0.7449 & 339.5191 & $-$34.2676 \\ 
22 & 3.92 & 739.6568 & $-$0.0921 & 313.2296 & $-$60.2544 \\ 
23 & 3.69 & 821.2659 & $-$1.2319 &  221.0008 & 27.1607 \\ 
24 & 3.62 & 825.1451 & $-$1.1582 & 74.7772 & $-$34.8971 \\ 
25 & 3.96 & 916.2110 &  1.6899 &  239.5039 & 27.3825 \\ 
26 & 3.47 & 949.4410 & $-$1.6215 & 155.2317 & $-$28.9975 \\ 
\hline  
\end{tabular} 
\caption[]{ List of sub-threshold candidates in the \fh\ analysis. Reference time is MJD 57856.826840. All parameters are averages of the values in each detector. } 
\label{tab:st}
\end{center} 
 \end{table*}

\subsection{{\Fstat} outliers} 

Table~\ref{tab:TDFstat_outliers} presents the parameters of the final 23 outliers
from the {\td}  pipeline, along with comments on their likely causes.
None is a credible gravitational wave signal.

\begin{table*}[tbp]
\begin{center}
\begin{tabular}{r r r r r r r}
\hline
\multicolumn{1}{c}{Index} & \multicolumn{1}{c}{FAP} & \multicolumn{1}{c}{Frequency} & \multicolumn{1}{c}{Spin-down} &  \multicolumn{1}{c}{$\alpha$}  & \multicolumn{1}{c}{$\delta$} & Description \\
\multicolumn{1}{c}{}  & \multicolumn{1}{c}{}  & \multicolumn{1}{c}{[Hz]}  &  \multicolumn{1}{c}{[nHz/s]} & \multicolumn{1}{c}{[deg]} & \multicolumn{1}{c}{[deg]} &  \\
\hline \hline
1 & $2.4\times10^{-4}$ & 21.428 &  0.014 & -63.801 & 100.39 & Present only in H1   \\ 
2 & $1.1\times10^{-4}$ & 27.481 &  0.027 & -63.159 & 272.42 & Present only in L1   \\ 
3 & $1.5\times10^{-5}$ & 29.970 & -0.003 & -66.303 & 89.028 & Present only in H1   \\  
4 & $1.3\times10^{-5}$ & 31.764 &  0.030 & -54.723 & 111.99 & Interference much stronger in H1 than in L1  \\  
5 & $6.3\times10^{-5}$ & 39.763 &  0.015 & -63.094 & 103.49 & Interference much stronger in H1 than in L1  \\  
6 & $3.2\times10^{-4}$ & 42.945 &  0.015 & -65.709 & 94.921 & Present only in H1     \\ 
7 & $4.5\times10^{-5}$ & 59.951 & -0.026 & -64.569 & 86.213 & Present only in H1     \\
8 & $< 10^{-8}$      & 82.753 & -4.229 & -52.811 & 294.54 & Interference stronger in H1 than in L1    \\
9 & $1.3\times10^{-4}$ & 83.447 &  0.011 & -65.403 & 94.618 & Interference much stronger in H1 than in L1    \\   
10 & $2.6\times10^{-4}$ & 85.714 & -0.003 & -65.907 & 90.414 & Present only in H1         \\ 
11 & $1.6\times10^{-4}$ & 85.830 & -0.004 & -63.420 & 88.385 & Present only in H1         \\ 
12 & $< 10^{-8}$      & 107.11 &  0.195 & -6.0084 & 209.42 & Present only in H1         \\  
13 & $< 10^{-8}$                & 107.14 &  0.032 & -6.4358 & 9.2631 &  Present only in H1        \\            
14 & $3.8\times10^{-5}$ & 119.90 &  0.046 & -6.8506 & 9.3225 &  Interference stronger in H1 than in L1  \\ 
15 & $< 10^{-8}$                & 128.57 &  0.362 & -7.0981 & 155.81 &  Present only in H1        \\  
16 & $< 10^{-8}$                & 130.93 &  0.122 & -5.2438 & 9.0533 &  Present only in H1        \\ 
17 & $7.9\times10^{-8}$ & 199.86 & -0.041 & -8.1570 & 138.45 &  Present only in L1        \\  
18 & $1.6\times10^{-8}$ & 299.41 &  0.132 & -8.2426 & 135.07 &  Present only in H1        \\            
19 & $< 10^{-8}$                & 299.54 &  0.331 & -7.0534 & 109.15 &  Present only in H1       \\ 
20 & $< 10^{-8}$                & 299.71 & -0.229 & -6.1941 & 7.7833 &  Present only in H1       \\ 
21 & $1.3\times10^{-7}$ & 303.27 &  0.034 & -3.4995 & 235.37 &  Present only in H1       \\  
22 & $< 10^{-8}$                & 314.91 & -0.145 & -6.7862 & 7.9226 &  Present only in L1       \\
23 & $< 10^{-8}$                & 1800.1 & -1.700 & -68.404 & 75.855 &  Present only in L1       \\
\hline
\end{tabular}
\caption{{\td} pipeline outliers in the frequencies ranges
$[20$-$100]$~Hz,  $[100$-$434]$~Hz, and  $[1518$-$1922]$~Hz.
  The columns provide outliers false alarm probability (FAP) as well as the nominal frequencies and frequency derivatives, right ascensions and declinations
  found for the outliers, along with comments indicating the likely sources of the outliers. Frequencies are converted to epoch GPS 1174899130.}
\label{tab:TDFstat_outliers}
\end{center}
\end{table*}

\begin{table*}[tbp]
\begin{center}
\begin{tabular}{r r r r r r}
\hline
Label & FA & Frequency [Hz] & Spin-down [nHz/s] & $\alpha$ [deg] & $\delta$ [deg] \\
\hline \hline
ip0 & $< 10^{-8}$ & 265.5746 (0.0007) &   $-$0.0466 ($-$0.0425) &  69.42 (2.12)  & $-$57.21 (0.99) \\
ip3 & $< 10^{-8}$ & 108.8573 ($-$0.0001) & $-$0.1386 (0.1386) &  174.96 (3.42) & $-$34.81 ($-$1.37) \\
ip5 & $2.9\times 10^{-4}$  & 52.8085 (0.0002) & $-$0.1865 (0.1865) &  236.90 (65.72) & $-$74.18 (9.66) \\
ip6 & $< 10^{-8}$ & 145.8721 (0.0048) & $-$6.0512 (0.6788) & 358.75 (58.00) & $-$68.01 (2.58) \\
ip8 & $< 10^{-8}$ & 190.6428 (0.0002) & $-$8.5306 (0.1193) & 331.41 (19.98) & $-$35.87 (2.45) \\
ip10 & $2.2\times 10^{-4}$  & 26.3380 (0.0001) & $-$0.0683 (0.0167) & 223.10 (1.54) & 37.23 (5.64)  \\
ip11 & $4.0\times 10^{-4}$  & 31.42475 (0.00001) & $-$0.0056 (0.0051) & 286.25 (1.10) & $-$58.39 (0.11) \\
ip12 & $5.7\times 10^{-2}$  & 38.2005 (0.0001) & $-$6.1165 (0.1335) & 326.26 (5.60) & $-$33.71 (16.73) \\
\hline
\end{tabular}
\caption[]{Hardware injection recovery with the \td\ pipeline. The values in parentheses are the absolute errors, that is, the difference with respect to the injection parameters. Frequencies are converted to epoch GPS 1174899130.}
\label{tab:tdfstat_hi}
\end{center}
\end{table*}

\end{document}